\documentclass[prb,floatfix,showpacs,superscriptaddress,floatfix,amsfonts]{revtex4}
\pdfoutput=1
\usepackage{color}
\usepackage[pdftex]{graphicx,hyperref}
\usepackage{dcolumn}
\usepackage{bm}
\usepackage{marvosym}
\usepackage{times,amsmath,amssymb}
\usepackage{array}
\usepackage{float}

\def\red#1{\textcolor[rgb]{1.00,0.00,1.00}{#1}}

\begin{document}

\title{Exciton-polaritons of a 2D semiconductor layer in a cylindrical microcavity}

\author{Jos\'e Nuno S. Gomes}
\affiliation{Centro de F\'{\i}sica, Universidade do Minho, Campus de Gualtar, Braga
4710-057, Portugal}
\author{Carlos Trallero-Giner}
\affiliation{Facultad de F\'{\i}sica, Universidad de La Habana, Vedado 10400, La Habana,
Cuba}
\affiliation{CLAF - Centro Latino-Americano de F\'{\i}sica, Avenida Venceslau Braz, 71,
Fundos, 22290-140, Rio de Janeiro, RJ, Brasil}
\author{Nuno M. R. Peres}
\affiliation{Centro de F\'{\i}sica, Universidade do Minho, Campus de Gualtar, Braga
4710-057, Portugal}
\affiliation{International Iberian Nanotechnology Laboratory, Braga, Portugal}
\author{Mikhail I. Vasilevskiy}
\affiliation{Centro de F\'{\i}sica, Universidade do Minho, Campus de Gualtar, Braga
4710-057, Portugal}
\affiliation{International Iberian Nanotechnology Laboratory, Braga, Portugal}

\begin{abstract}

We describe exciton-polariton modes formed by the interaction between excitons in a 2D layer of a  transition metal dichalcogenide  embedded in a cylindrical microcavity and the microcavity photons. For this, an expression for the excitonic susceptibility of a semiconductor disk placed in the symmetry plane perpendicular to the axis of the microcavity is derived.
Semiclassical theory provides dispersion relations for the polariton modes, while the quantum-mechanical treatment of a simplified model yields the Hopfield coefficients, measuring the degree of exciton-photon mixing in the coupled modes.
The density of states (DOS) and its projection onto the photonic and the excitonic subspaces are calculated taking monolayer MoS$_2$ embedded in a Si$_3$N$_4$ cylinder as an example.
The calculated results demonstrate a strong enhancement, for certain frequencies, of the total and local DOS (Purcell effect) caused by the presence of the 2D layer.

\end{abstract}
\maketitle

\section{Introduction}
Strong light-matter coupling takes place in nearly two-dimensional (2D) semiconductors of the transition metal dichalcogenide (TMD) family, making these materials highly interesting for a range of photoelectronic applications including photodetection and lasing, as well as from the fundamental physics point of view.~\cite{Britnell2013,Liu2015,Koperski2015} Excitons in TMDs are very robust, with the binding energy of the order of 0.5 eV and a very small effective Bohr radius,~\cite{Wang2018} which imply the presence of bound excitons at room temperature and above and the possibility of 3D confinement of such an exciton as a whole. Together with the rather strong dipole optical transition taking place in the $K$ and $K^\prime$ points of the Brillouin zone, highly anisotropic emission owing to the 2D nature of the underlying electronic states, and specific selection rules with (circular-) polarization-valley correlation, these materials' characteristics are unique indeed. Compared to traditional semiconductor quantum wells, mono- or few-layer TMDs can more easily be combined with other 2D materials such as graphene and h-BN to make van der Waals heterostructures,~\cite{Geim2013} as well as be coupled to other quantum emitters~\cite{Cartstein2015,Karanikolas2016,Raja2016} or plasmonic nanostructures.~\cite{Najmaei2014,Liu2016,Abid2017,Liu2018}. Recently, integration of an electroluminescent van der Walls heterostructure containing a TMD layer cladded by h-BN and graphene layers into a monolithic optical micro-cavity was demonstrated, leading to a strong increase and a modification of the angular distribution of the electrically pumped emission from the TMD layer.~\cite{Tartakovskii2019}

Coupling between the  light and excitons can be greatly enhanced by placing the semiconductor structure into a microcavity (MC) where the photons are confined. In the strong coupling regime, the exchange of energy between excitons and photons becomes reversible, yielding a number of interesting and potentially useful effects owing to the formation of collective excitations called exciton-polaritons.~\cite {Kavokin_MCs,Andreani2014} Modification of the energy spectra of the system because of the strong light-matter interaction has been achieved for a variety of composite structures that may be generally named MCs and include Fabry-Perot cavities with Bragg reflectors as mirrors, micropillars, photonic crystals and plasmonic surfaces and nanostructures (see Ref.~\onlinecite{Dovzhenko2018} for a recent review). The possibility to realize the strong coupling regime between {\it 2D excitons} and MC photons has been analyzed theoretically and demonstrated experimentally, for usual Fabry-Perot~\cite{Liu2015,Dufferwiel2015,Vasilevskiy2015,Flatten2016} and the so called Tamm-type MCs,~\cite{Lundt2016} and also for a cylindrical whispering gallery mode (WGM) resonator.~\cite{Ye2015} Integrating van der Waals heterostructures with optical waveguides has recently been reviewed.~\cite{C-H_Liu2019}
The encouraging results include high Rabi splittings (comparable to the best GaAs and II-VI MCs), the existence of exciton-polaritons at room temperature, and 2D-exciton-mediated lasing. The latter was achieved with a WGM microdisk resonator containing an embedded WS$_2$ monolayer.~\cite{Ye2015} Such a closed geometry ensures better confinement of light (smaller mode volume) and, potentially, can provide a stronger exciton-photon coupling compared to the traditional Fabry-Perot type of cavity. In this work, we perform a theoretical analysis of polaritons formed by 2D excitons in a TMD layer placed in the symmetry plane (perpendicular to the axis) of a cylindrical cavity and the confined MC photons.

\begin{figure}[H]
\includegraphics[width=.3\linewidth]{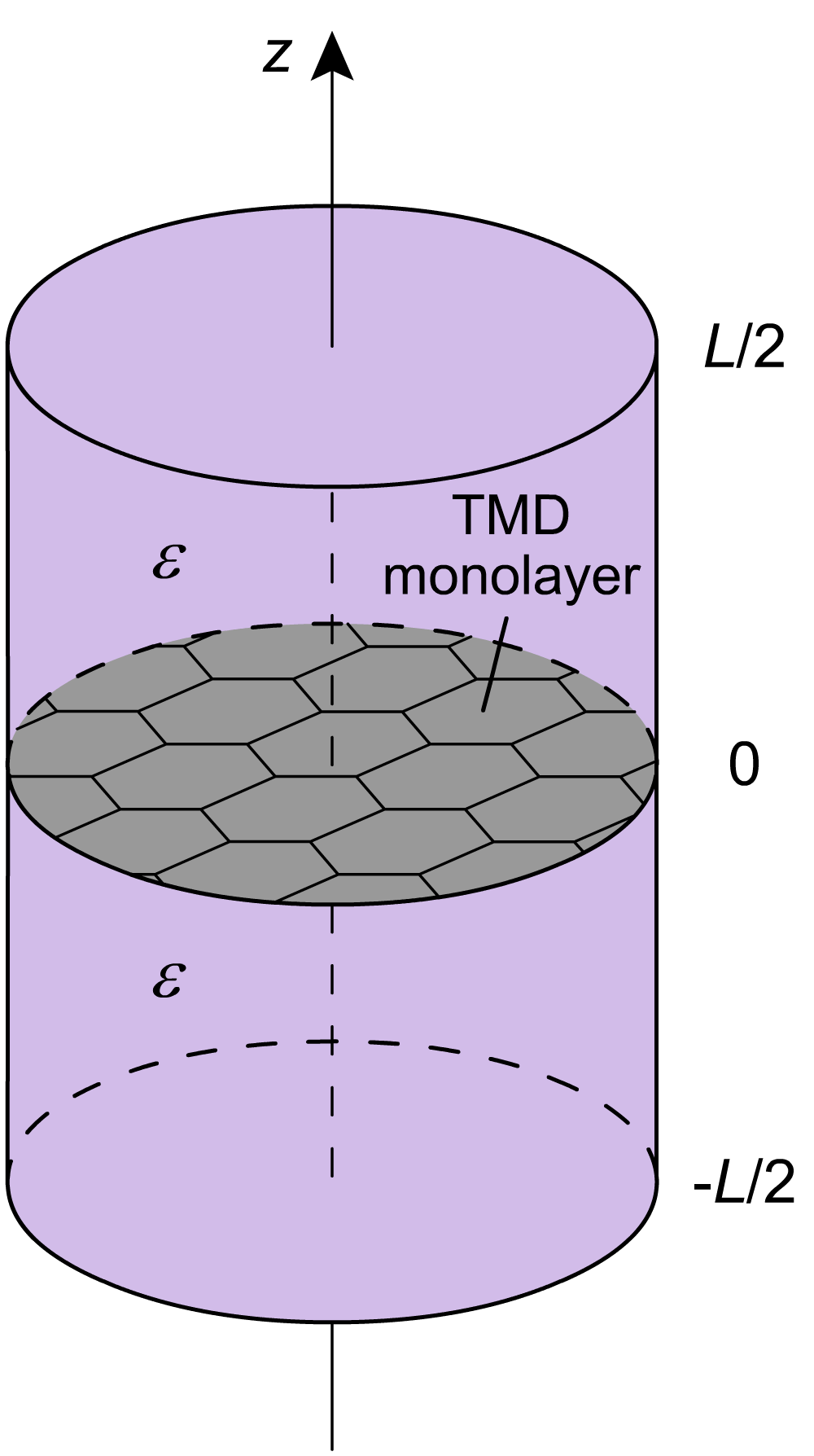}
\centering
\caption{(Color online) Sketch of the system under study:  a TMD monolayer inserted in a dielectric cylindrical cavity of length $L$ with relative permittivity $\epsilon $.}
\label{scheme}
\end{figure}

The system under consideration is depicted in Fig.~\ref{scheme}.
As known, Maxwell's equations cannot be solved for a cylinder of finite height ($L$) embedded in an infinite medium with a different (finite) dielectric constant.~\cite{Jackson} Since our goal is to tackle the problem mostly analytically, we shall assume that the electromagnetic (EM) fields are perfectly confined inside the cylinder as if it were embedded in a perfect metal. The validity of this approximation will be examined in the $L\rightarrow \infty$ case. Within the perfect EM confinement model, we shall present calculated exciton-polariton modes (density of states and the dispersion relation), the Hopfield coefficients measuring the fraction of exciton and photon in a polariton mode,~\cite{Hopfield} and the emission enhancement factor for a point emitter located in the TMD plane (e.g. a point defect). For such an emitter, we shall assume a weak coupling regime and analyze how its emission is enhanced or inhibited (the Purcell effect~\cite{Purcell1946}) due to the polaritonic background of the microcavity with embedded TMD layer. 
%
%
The article is organized as follows. In the next section we consider excitons confined to a 2D disk and derive the electric susceptibility of the TMD layer, which determines the coupling effects of the excitons with the electromagnetic field of the cavity modes. In Sec. III we study coupled (polariton) modes in such a cavity using both semiclassical and quantum approaches. Employing these results, in Sec. IV we  evaluate the total and projected densities of states (DOS) showing the possibility of strong enhancement of the local photonic DOS modified due to the presence of the TMD layer in the cavity. In addition, we provide a theoretical analysis of the Purcell factor of a point emitter located in the vicinity of the TMD layer. The last section is devoted to discussion and conclusions and the reliability of the perfect EM confinement approximation is analyzed in Appendices~\ref{appendix A} and  \ref{appendix B}.

\section{Excitonic susceptibility of a 2D semiconductor embedded in a cylindrical cavity}
\subsection{Exciton confinement}

Let us consider tightly bound excitons in a 2D semiconductor disk of radius $R$.
We notice that the exciton's effective Bohr radius, $a_{ex}$, is very small, of the order of 1 nm,~\cite{Wang2018} i.e. $a_{ex}\ll R$.
This allows us to disregard the exciton's internal structure in what concerns its confinement in the radial direction and treat the exciton as a point-like particle, which is simply confined in a circular potential well of radius $R$. The potential acting on the exciton centre of mass outside of the disk can be considered infinite. The Schr\"odinger equation for the centre of mass of the exciton takes the form:
\begin{equation}
-\frac{\hbar^2}{2M_{ex}}\nabla_{2D}^2 \Psi_{ex}(\vec{r})=(E-E_{ex}) \Psi_{ex}(\vec{r})\;,
\label{eq.exciton}
\end{equation}
where $\vec{r}$ is a 2D radius-vector, $M_{ex}$ stands for the excitonic mass and $E_{ex}$ is its binding energy for an infinite 2D layer cladded by the MC dielectric material.
The solution of Eq.~(\ref{eq.exciton}) in polar coordinates can be cast as
\begin{equation}
\Psi_{\pm m,n}(\vec{r})=C_{m,n} J_m(q_{m,n} r){\large e}^{\pm im\phi}\;,
\label{Psi_mn}
\end{equation}
being $m\ge 0$, $n\ge 1$ the angular momentum and radial quantum numbers, respectively, $q_{m,n}=\xi_{m,n}/R$ the radial wave-number, $\xi_{m,n}$ the $n$'th zero of the Bessel function of the first kind, $J_m(x)$, and
$C_{m,n}=1/(\sqrt{\pi}R\vert J_{m}^\prime(\xi_{m,n})\vert)$ is a normalization constant.
The exciton energy $E\mapsto E_{m,n}$ is given by
\begin{equation}
 E_{m,n}=E_{ex}+\frac{\hbar ^2 \xi_{m,n}^2}{2M_{ex}R^2}\;.
\label{eq.exc-energy}
\end{equation}

\subsection{Photon modes in empty cylindrical cavity}
Confined photon modes of the empty cavity obey the Maxwell's equations and can be classified by the $z$-component of the photon wavevector, $k_z$, and two indices, $\mu$ and $\nu$, that take account the axial symmetry and the radial photon confinement, respectively (the last two determine the in-plane component of the wavevector and are analogous to the numbers $m$ and $n$ defined for excitons confined to a disk). In order to make our consideration simple, we shall assume  the approximation of ``ideal cavity'' with perfectly reflecting walls, which has been used by many authors (e.g. Ref.~\onlinecite{Gutbrod1998}). A more eleborate treatment of photonic eigenmodes in a cylindrical microcavity considering a finite dielectric constant outside of the cavity  (and, consequently, imperfect confinement of the fields inside it) has been presented in Ref. ~\onlinecite {Panzarini1999} using approximate decoupling the degrees of freedom along and perpindicular to $z$ direction and the cavity modes are evaluated as a function of radius according to a self-consistent procedure. The approximation of perfectly reflecting walls has the obvious advantage of obtaining the eigenmodes analytically.    
The reasonable validity of this approximation is illustrated in Appendix~\ref{appendix A} where the field profiles are compared with the exact solution, in the case of a very long cylinder ($L \rightarrow \infty $); the eigenmode frequencies also are not considerably different for the range of parameters we are interested in here. 

In the framework of this approximation, we can distinguish purely transverse electric (TE) and transverse magnetic (TM) modes, which yields an additional label representing mode polarization. Below we shall limit ourselves by considering only TM modes.
For a perfectly confined TM mode we have the following field components
\begin{eqnarray}
E_z(r,\phi,z,\omega)=E_0 J_\mu(q_{\mu,\nu} r) \cos\bigg[k_z\big(\frac{L}{2}-z\big)\bigg] e^{i(\mu\phi-\omega t)}\;,
\label {E_z}\\
E_r(r,\phi,z,\omega)=E_0J_{\mu}'(q_{\mu,\nu}r)\frac{k_z}{q_{\mu,\nu}} \sin\bigg[k_z\big(\frac{L}{2}-z\big)\bigg]e^{i(\mu\phi-\omega t)}\;,
\label {E_r}\\
E_{\phi}(r,\phi,z,\omega)=i E_0 \mu J_\mu(q_{\mu,\nu} r) \frac{k_z}{q_{\mu,\nu}^2 r} \sin\bigg[k_z\big(\frac{L}{2}-z\big)\bigg] e^{i(\mu\phi-\omega t)} \;,
\label {E_phi}\\
H_r(r,\phi,z,\omega)=E_0 \mu \epsilon J_\mu(q_{\mu,\nu} r) \frac{\omega}{cq_{\mu,\nu}^2 r} \cos\bigg[k_z\big(\frac{L}{2}-z\big)\bigg]  e^{i(\mu\phi-\omega t)}\;,
\label {H_r}\\
H_{\phi}(r,\phi,z,\omega)=i E_0 \epsilon J'_\mu(q_{\mu,\nu} r)\frac{\omega}{cq_{\mu,\nu}} \cos\bigg[k_z\big(\frac{L}{2}-z\big)\bigg] e^{i(\mu\phi-\omega t)}  \;,
\label {H_phi}
\end{eqnarray}
where $q_{\mu,\nu}=\xi_{\mu,\nu}/{R}$ now stands for the radial part of the photonic wavenumber,
and
\begin{equation}
k=\sqrt{k_z^2+q_{\mu,\nu}^2} = \sqrt{\epsilon}\frac{\omega}{c}
\label{k-vector}
\end{equation}
is the total wavenumber of the photon.
Consequently, the eigenfrequencies of the empty cavity modes are given by the relation
\begin{equation}
\omega (\mu,\nu, k_z)=\frac c {\sqrt{\epsilon}} \sqrt{k_z^2+q_{\mu,\nu}^2}\;,
\label{eq.photon-energy}
\end{equation}
where the polarization index was skipped for clarity. For $L \rightarrow \infty $ (a waveguide), $k_z$ can be arbitrary, while for a cavity with perfectly reflecting bases
$k_z=\pi l/L \; (l=1,2,\dots ).$
In order to quantize this field for a single photon, we need to determine the amplitude $E_0$ that corresponds to the energy normalization condition~\cite{Berestetskii_QE}
\begin{equation}
\hbar \omega = \frac{1}{4\pi}\int d^3r\big(\epsilon \vec{\mathbb{E}} \cdot \vec{\mathbb{E}}^* + \vec{\mathbb{H}} \cdot \vec{\mathbb{H}}^*\big)\;.
\label{energycons}
\end{equation}
It is convenient to define the so called mode volume, $\Omega (\mu,\nu, k_z)$, and write down the photon field amplitude in the form analogous to that for a photon in vacuum~\cite{Berestetskii_QE}
\begin{equation}
E_0(\mu,\nu,k_z)=\;\sqrt{\frac{2\pi\hbar\omega (\mu,\nu, k_z)}{\epsilon \Omega (\mu,\nu, k_z)}}\;,
\label{E0}
\end{equation}
with
\begin{equation}
\Omega (\mu,\nu, k_z)=\frac{1}{2\epsilon}\int d^3 {r} \big(\epsilon \frac{\vec{\mathbb{E}} \cdot \vec{\mathbb{E}}^*}{E_0^2}+ \frac{\vec{\mathbb{H}} \cdot \vec{\mathbb{H}}^*}{E_0^2}\big)\;.
\label{modevolume}
\end{equation}
Its explicit expression obtained by substituting Eqs~(\ref{E_z})-(\ref{H_phi}) into (\ref {modevolume}) is:
\begin{equation}
\Omega (\mu,\nu,k_z)=\frac{\pi R^2}{2k_z}\bigg[J_{\mu+1}(\xi_{\mu,\nu}))\bigg]^2\bigg[k_zL+\frac 12 \sin(2k_zL)+\frac{L R^2k_{z}^3}{\xi_{\mu,\nu}^{2}}\bigg]
\;.
\end{equation}
For an ideal cavity, $k_z=\pi l/L$, $\Omega (\mu,\nu,k_z)\mapsto\Omega (\mu,\nu,l)$ and
\begin{equation}
\Omega (\mu,\nu,l)=\frac{\pi R^2 L}{2}\bigg[J_{\mu+1}(\xi_{\mu,\nu}))\bigg]^2\bigg[1+\frac{\pi^2 l^2R^2}{\xi_{\mu,\nu}^2L^{2}}\bigg]
\;.
\label{modevolume2}
\end{equation}
Its dependence upon the mode energy is shown in Fig. \ref{Fig:modevolume} in Appendix~\ref{appendix A}. As expected, it is much smaller than for planar cavities owing to the field confinement in all directions.

\subsection{Susceptibility and optical conductivity}
The exciton-photon interaction, in the dipole approximation is given by
\begin{equation}
{V}_{int}=-(\vec{d}_{CV} \cdot \vec{\mathbb{E}})\;,
\label {V_int}
\end{equation}
where $\vec{d}_{CV}$ is the dipole moment matrix element between the valence and conduction bands,~\cite{DiXiao2012}
\begin{equation}
\vec{d}_{CV}=\frac {{d}_{CV}}{\sqrt{2}}(\vec e_x \pm i\vec e_y)\;;\qquad
{d}_{CV} = \frac{v_F \hbar e}{E_{ex}}
\label {d_cv} \; ,
\end{equation}
where $v_F$ is the Fermi velocity and $\pm$ applies to $K$ or $K^\prime$ points of the Brillouin zone.~\cite{McDonald2015} In Eq.~(\ref {V_int}), $\vec{\mathbb{E}}$ is the electric field associated with the empty cavity photon.
With the interaction operatot~(\ref {V_int}), we can write the matrix element for the process of creation of an exciton in a state $(m,n)$ at the cost of one photon $(\mu,\nu, k_z)$ as follows\red{:}
\begin{equation}
\mathcal{M}(\mu,\nu, k_z;m,n)=\langle m,n \vert -\vec{d}_{CV} \cdot \vec{\mathbb{E}}(\mu,\nu, k_z ) \vert  0\rangle \;,
\label {ME1}
\end{equation}
 where $ \vert  0\rangle$ stands for exciton vacuum,
\begin{equation}
\nonumber
\vert  0\rangle = \delta (\vec{r}_e-\vec{r}_h)\;,
\end{equation}
with $\vec{r}_e\, (\vec{r}_h)$ denoting electron (hole) radius-vector in the 2D TDM disk (see Fig.~\ref{scheme})
and the excitonic state $\vert m,n \rangle$ can be written as
\begin{equation}
\vert m,n \rangle = \Phi (\vec{r}_e-\vec{r}_h) \Psi _{m,n}(\vec{r})\;.
\label{excwf}
\end{equation}
Here, $\Phi (\vec{r}_e-\vec{r}_h)$ is the part of the excitonic wavefunction that represents its internal structure. Its magnitude in the origin can be evaluated using the hydrogen-like model~\cite{Ciuti1998}
\begin{equation}
\vert \Phi(0) \vert ^2=\frac{2}{\pi a_{ex}^2}\;,
\label{excwf0}
\end{equation}
while the centre-of-mass part is given by Eq.~(\ref {eq.exc-energy}).
\cite{Nota1}

Using Eq.~(\ref {d_cv}), the expressions for the electric field components and the known relations for the Bessel functions, Eqs.~(\ref {E_z}) and (\ref {E_r}), we obtain:
\begin{equation}
\nonumber
\big(\vec{d}_{CV} \cdot \vec{\mathbb{E}}(\mu,\nu, k_z )\big)\vert _{z=0}=
\frac {k_zR}{\sqrt 2 \xi_{\mu,\nu}}d_{CV}E_0 \sin \left (k_z {L}/{2} \right )
\left [e^{i(\mu - 1)\phi}J_{\mu - 1}(\xi_{\mu,\nu} r/R) - e^{i(\mu +1)\phi}J_{\mu +1}(\xi_{\mu,\nu} r/R)\right ]\;.
\end{equation}
With this relation, we get for the matrix element~(\ref {ME1}):
 \begin{equation}
\mathcal{M}(\mu,\nu, k_z;m,n)={d}_{CV}\Phi(0) \frac {k_zS} {\sqrt {2 \pi}} \sin \left (k_z {L}/{2} \right )\sqrt{\frac{2\pi\hbar\omega (\mu,\nu, k_z)}{\epsilon \Omega (\mu,\nu, k_z)}}I_{\mu,\nu;m,n}\left ( \delta _{\mu + 1, m} - \delta _{\mu - 1, m} \right )\;,
\label {ME2}
\end{equation}
where $S=\pi R^2$ and the coefficients $I_{\mu,\nu;m,n}$ are defined in Appendix~\ref{appendix B}.

The exciton-photon coupling takes place only for modes which are odd with respect to $E_z$ (otherwise $ \sin \left (k_z {L}/{2} \right ) =0$).
The Kronecker symbols in~(\ref  {ME2}) express the angular momentum conservation but there is no such property with respect to the radial numbers $\nu$ and $n$.
The dipole moment vector lies in the plane of the TMD layer, so the susceptibility tensor has two equal components ($xx$ and $yy$), while the third one vanishes. It is convenient to consider two-dimensional (sheet) susceptibility defined as the dipole moment per unit area per unit field. This susceptibility can be obtained using the second order perturbation theory similar to the derivation of the atomic polarizability.~\cite {LL-III} We write the energy of the semiconductor disk polarized by an electromagnetic mode $(\mu,\nu,k_z)$ as
\begin{equation}
\text E =\frac 1 2 \chi_{2D} \int_{disk} ( \vec{\mathbb{E}}_\bot  \cdot  \vec{\mathbb{E}}_\bot ^\star ) \vert _{z=0} d\vec r =\sum _{m,n}\frac{\vert \mathcal{M}(\mu,\nu, k_z;m,n) \vert ^2}{E_{m,n}-\hbar \omega}\;,
\end{equation}
where ${\vec{\mathbb{E}}}_\bot$ is the in-plane component of the electric field [see Eqs.~(\ref {E_r}) and (\ref {E_phi})].

Therefore, we obtain for the susceptibility:
\begin{equation}
\chi_{2D} =\frac 2 {SB_{\mu,\nu} \sin ^2 \left (k_z {L}/{2}\right )}
\sum_{m,n} \frac{\vert \mathcal{M}(\mu,\nu, k_z;m,n) \vert ^2}{E_{m,n}-\hbar \omega - i\delta }\;,
\label{susceptibility}
\end{equation}
where a small imaginary part $ i\delta$  has been added to the photon energy as usual and
$$
B_{\mu,\nu}=\vert E_0 \vert ^2 \left (\frac {k_z R}{\xi _{\mu,\nu}} \right )^2 \int_0^1{J_{\mu+1}^2(\xi_{\mu,\nu}x)xdx}=\vert E_0 \vert ^2 \left (\frac {k_z R}{\xi _{\mu,\nu}} \right )^2 \left [J^{\prime}_{\mu}(\xi_{\mu,\nu}R)\right ]^2\;.
$$
Here we have used the known relations between the Bessel functions and their derivatives~\cite{Abramowitz} and the fact that  $J_{\mu}(\xi_{\mu,\nu}R)=0$ for the considered cavity modes. 
The expression (\ref {susceptibility}) neglects the non-resonant term that is obtained by replacing $\omega \rightarrow -\omega$, which follows from Green's function formalism.~\cite {Haug-Koch}

Substituting the matrix element  from Eq.~(\ref{ME2}), the formula for the susceptibility reduces to
\begin{equation}
\chi_{2D} (\omega ;\mu,\nu)=\frac {\left \vert {d}_{CV}\Phi(0) \right \vert ^2\xi_{\mu,\nu}^2}{\left [J^{\prime}_{\mu}(\xi_{\mu,\nu}R)\right ]^2}\sum _{m,n}\frac{\vert I_{\mu,\nu;m,n} \vert ^2}{E_{m,n}-\hbar \omega  - i\delta} \left ( \delta _{\mu + 1, m} + \delta _{\mu - 1, m} \right )\;,
\label{susceptibility2}
\end{equation}
where the integrals $I_{\mu,\nu;m,n}$ [defined in Eq. \eqref {ME2}] are presented in Appendix~\ref{appendix B}. 
Notice that the susceptibility does not depend explicitly on $k_z$ but does depend on $\mu $ and $\nu $, which determine the in-plane component of the photon wave-vector.
Equation~(\ref  {susceptibility2}) assumes only one type of exciton. It is known that spin-orbit interaction in TMDs results in two types of excitons, usually called $A$ and $B$ with a splitting of 100-200 meV.~\cite{Wang2018} Generalization of~(\ref  {susceptibility2}) to this situation is straightforward by adding a sum over $X=A,B$. However, in order to simplify the subsequent consideration, we shall include only lower energy ($A$) excitons.
For usage in the next section, it is convenient to define  the optical conductivity, $\sigma_{2D}(\omega)$, which is related to the 2D susceptibility by
\begin{equation}
\sigma_{2D} (\omega ;\mu,\nu))=-i\omega\chi _{2D}(\omega ;\mu,\nu)\;.
\end{equation}

\section{Microcavity exciton-polaritons}

\subsection{Semiclassical theory}
\begin{figure}[!htb]\centering
   \begin{minipage}{0.48\textwidth}
     {\includegraphics[width=.95\linewidth]{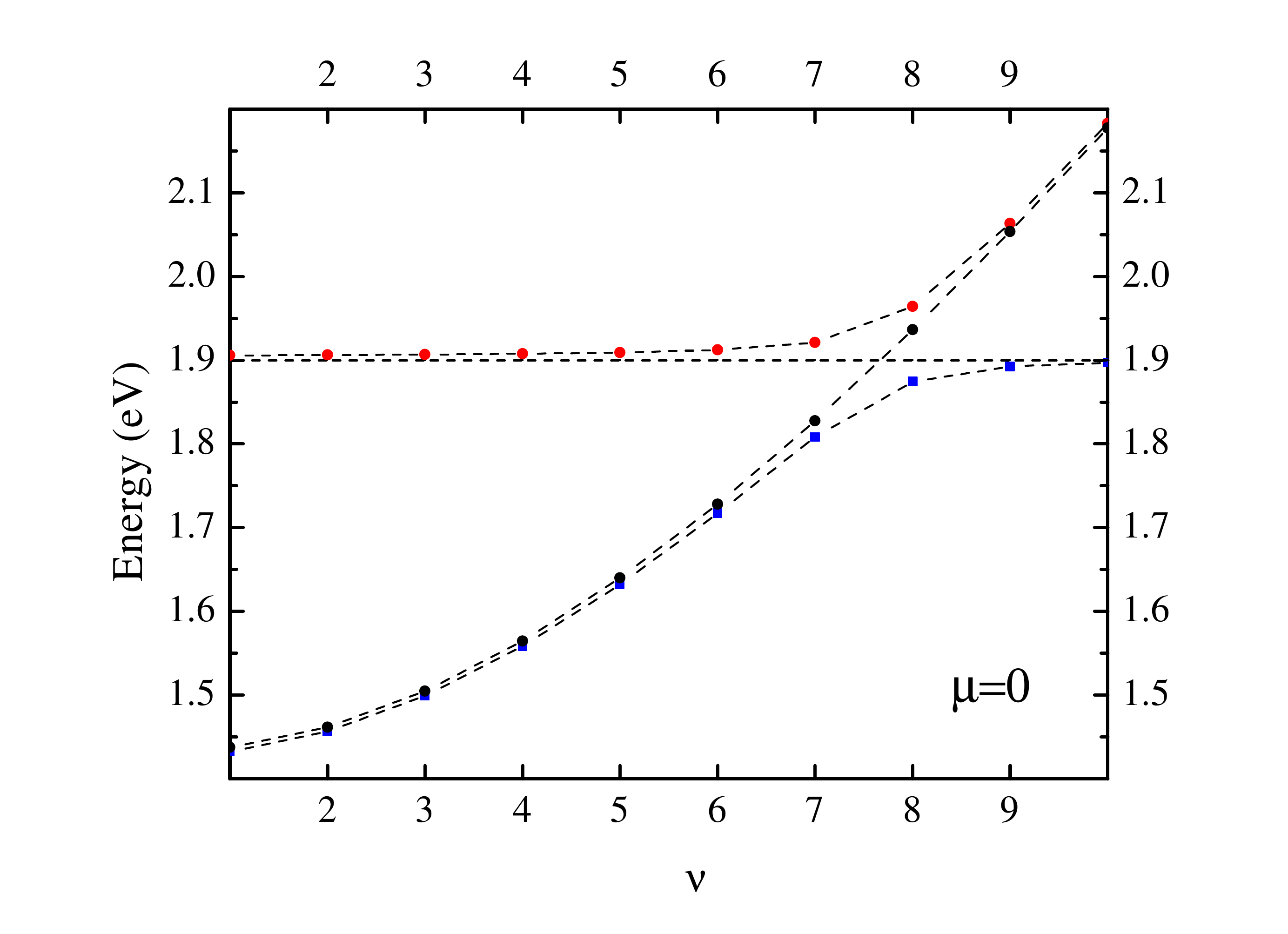}}
   \end{minipage}
   \begin {minipage}{0.48\textwidth}
     {\includegraphics[width=.95\linewidth]{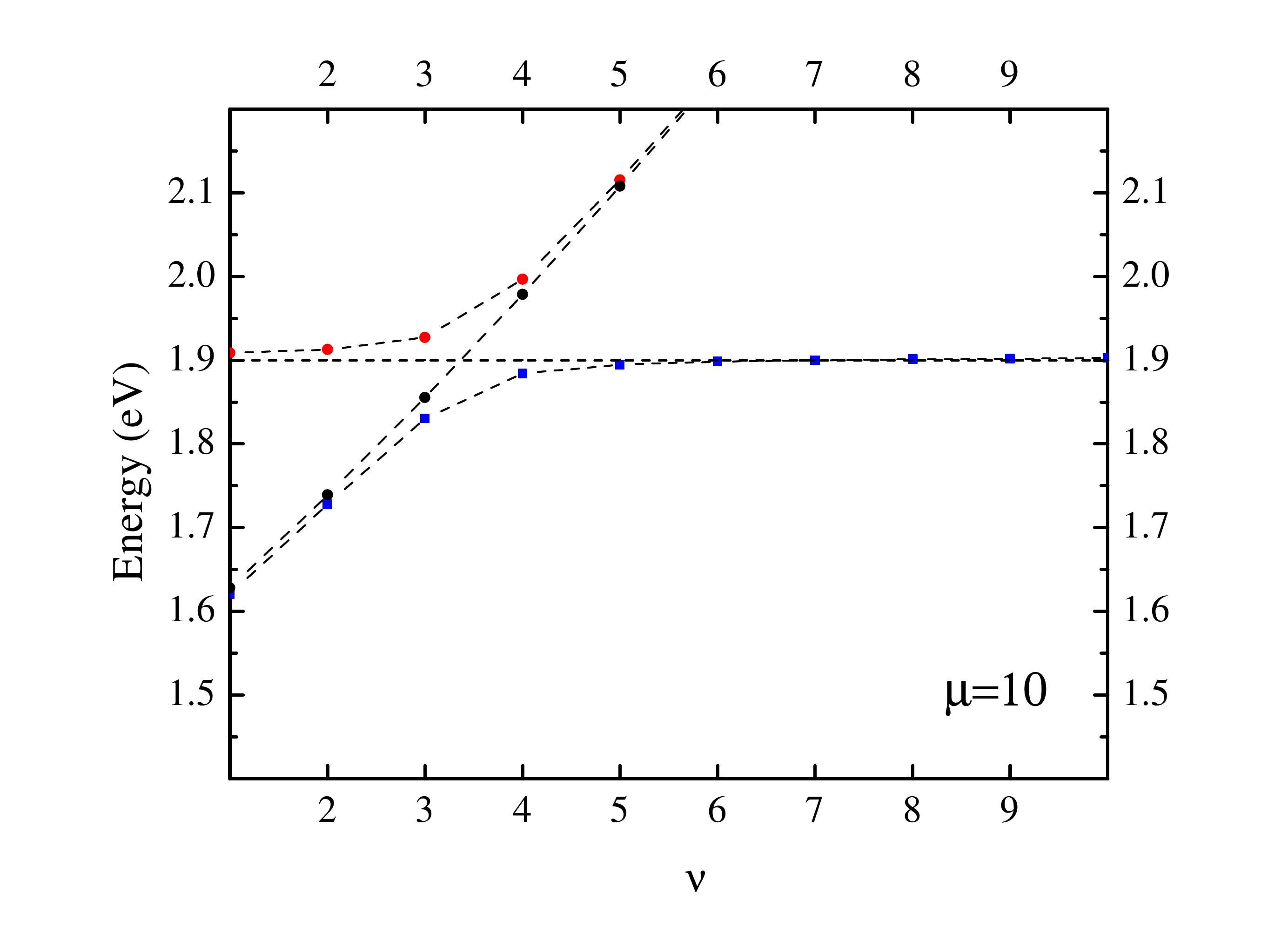}}
   \end{minipage}
\caption{(Color online) Uncoupled exciton (dashed line), MC photon (dashed line with black balls) and polariton dispersion curves for $\mu=0$  and 10 (red balls and blue squares, respectively). The radial index takes only integer values but the points are connected to improve visibility.
The parameters are: cylinder radius $R=2\;\mu$m, height $L=0.235\; \mu$m, dielectric constant of the MC material $\epsilon =3.4$, Fermi velocity $v_F=5.5\times10^5\; m/s$, exciton energy $E_{ex}=1.9\;eV$, and Bohr radius $a_{ex}=0.8 \;  nm$.
}
\label{Fig:classdispcurves}
\end{figure}

In order to determine the allowed states, given this interaction, it is necessary to plug in the following boundary conditions~\cite{c:primer}
\begin{eqnarray}
\nonumber
H^+_r-H^-_r=\frac{4\pi}{c}\sigma_{2D} E_{\phi}\vert _{z=0}\;,\\
H^+_{\phi}-H^-_{\phi}=-\frac{4\pi}{c}\sigma_{2D} E_r\vert _{z=0}\;.
\label{BCs}
\end{eqnarray}
Here the $+(-)$ sign stands for the  fields at $z\ge 0\;  (z\le 0)$.
We write the fields above and below the TMD layer in the form:
\begin{equation}
E_z^{\pm}(r,\phi,z,\omega)=\pm E_0 J_\mu(q_{\mu,\nu} r) \cos\bigg[k_z(\omega)\big(\frac{L}{2}\mp z\big)\bigg] e^{i(\mu\phi-\omega t)}
=-\frac{4\pi}{c}\sigma_{2D} E_r\vert _{z=0}\;.
\label{Epm}
\end{equation}
Using~(\ref{Epm}),~Eqs. (\ref{BCs}) yield the following polariton dispersion relation
\begin{equation}
{2\pi\sigma_{2D} (\omega ;\mu,\nu)}k_z \tan(k_z \frac{L}{2})=-i\omega \epsilon \,.
\label{mode-equation}
\end{equation}
This equation, together with Eq.~(\ref{k-vector}), permits to obtain the allowed values of $k_z$ (they are, of course, different from those of the empty cavity, but will be labelled by the same index $l$) and the corresponding $\omega $ for each pair of quantum numbers $\mu $ and $\nu$
\begin{equation}
\omega (\mu,\nu,l)=\frac{c}{\sqrt{\epsilon}}\sqrt { k_z^2(\mu,\nu,l) + \left (\frac{\xi_{\mu,\nu}}{R}\right)^2}\;.
\label{eigenfrequencies}
\end{equation}
The polariton "dispersion curves" for two angular momentum values are shown in Fig.~\ref{Fig:classdispcurves}, which presents the eigenfrequencies calculated by Eqs.~(\ref{mode-equation})  and  (\ref{eigenfrequencies}) {\it versus} the radial imdex, $\nu $, for the lowest $k_z$ ($l=1$). 
%
%
%
%
\begin{figure}[H]
\includegraphics[width=.5\linewidth]{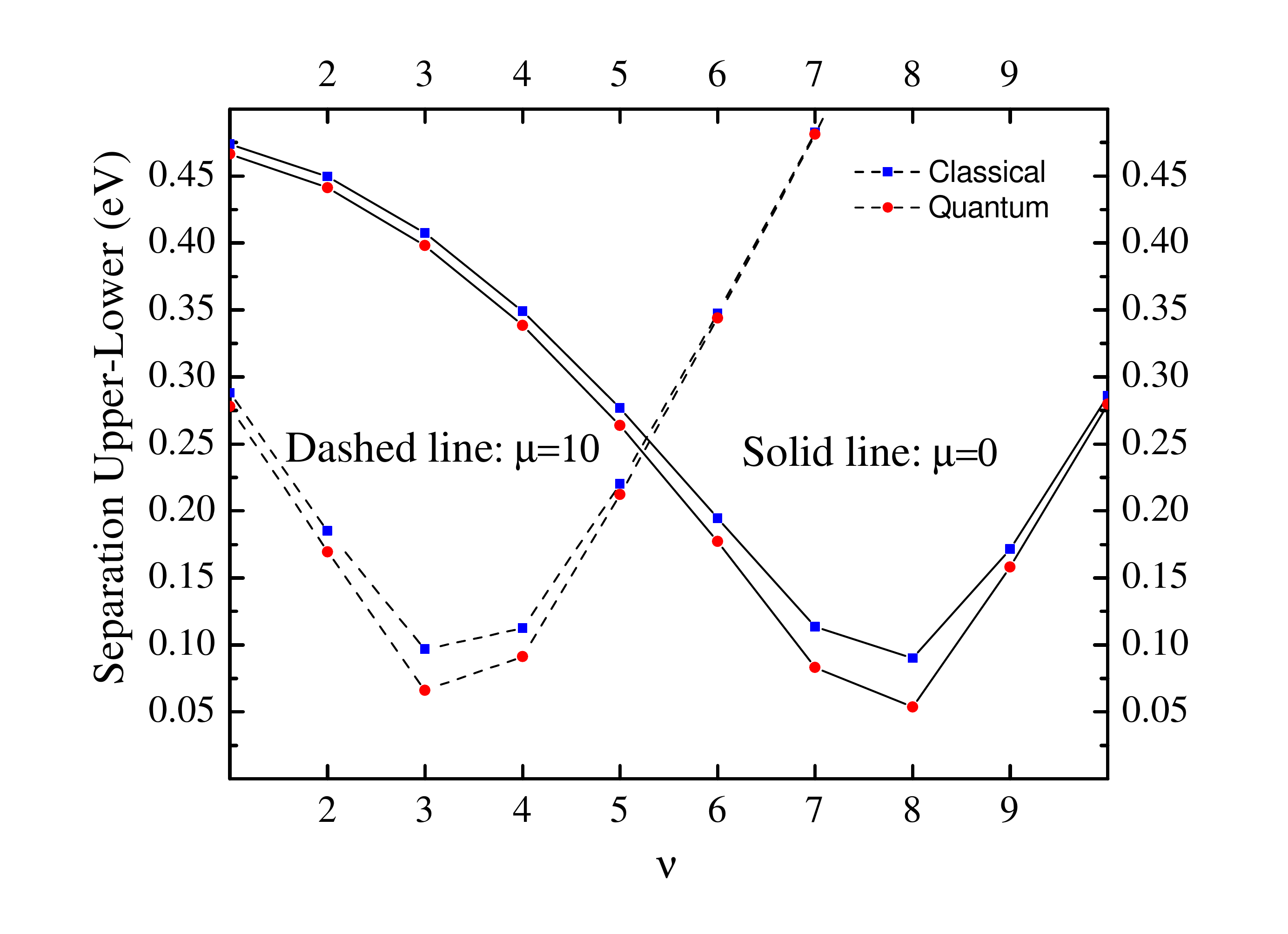}
\centering
\caption{(Color online) Separations between the upper and lower polariton modes calculated within the classical picture, corresponding to  Fig. \ref{Fig:classdispcurves}(red balls) and using the quantum formalism (blue squares), for $\mu=0$ (connected by solid lines)  and 10 (connected by dashed lines).
The parameters are the same as in Fig. \ref{Fig:classdispcurves}.
}
\label{Fig:separations}
\end{figure}

\subsection{Quantum picture}

Quantum-mechanical interaction between two bosonic fields can be considered by applying a unitary transformation proposed by Hopfield.~\cite{Hopfield} As in~Ref.~\onlinecite{Vasilevskiy2015}, we denote by  $P_{(\dots)}$ ($P_{(\dots)}^\dagger$)  and $A_{(\dots)}$ ($A_{(\dots)}^\dagger$)  the annihilation (creation) operators for photons and excitons, respectively, where $(\dots)$ stands for the appropriate quantum numbers. The interaction term in the Hamiltonian is a sum over all these quantum numbers, with the interaction constant, in our case, equal to the matrix element~(\ref {ME1}). In a Fabry-Perot MC, the in-plane quantum numbers are replaced by $m,n \rightarrow \vec q$ for excitons and  $\mu,\nu \rightarrow \vec k$ for photons. Because of the uniformity of the system in any direction perpendicular to $z$, the interaction matriz element in a planar microcavity contains the Kronecker symbol $\delta _{\vec {k},\,\vec {q}}$ . By virtue of this, the Hamiltonian of interacting excitons and MC photons, in the Fabry-Perot case reduces to the form~\cite{Vasilevskiy2015} $\hat{H}=\sum_{\vec q}  \hat{H}_{\vec q}$ with
\begin{equation}
\hat{H}_{\vec q}=\hbar\omega(\vec q)P_{\vec q}^{\dagger}P_{\vec q}+E_{ex}(\vec q)A_{\vec q}^{\dagger}A_{\vec q}+ g(\vec q)P_{\vec q}^{\dagger}A_{\vec q}+\text {H.C.}\;,
\label{hamFP}
\end{equation}
where $g(\vec q)$ is the coupling constant for the cavity mode $\vec q$ (for a certain $l$). Therefore, the Hopfield transformation can be applied to diagonalize this Hamiltonian, for each $\vec q$ separately.

Now, for the case of cylindrical cavity, the coupling constant  (for a certain $k_z$), $g({\mu,\nu,k_z;m,n})$ is given by Eq.~(\ref {ME1}) and we have only Kronecker deltas involving the quantum numbers $m$ and $\mu$ and there are no restrictions on $\nu$ and $n$. If we include all possible combinations of $\nu$ and $n$ in the Hamiltonian, we will not be able to write down a tractable Hopfield transformation. And even if we include only forms with $|\nu-n|=0,\pm1$ there will be seven Hopfield coefficients and the diagonalization will become impossible to perform analytically.  We  shall make the following (seemingly crude) approximation, $n=\nu$, which will be checked afterwards.
Therefore, we can write the Hamiltonian (for a certain $k_z$ which is omitted for clarity; in the following we shall consider the lowest branch with $l=1$) as
\begin{equation}
\hat{H}=\sum_{\mu,\nu}\bigg[\hbar \omega (\mu,\nu)P_{\mu,\nu}^{\dagger}P_{\mu,\nu}+E_{ex}A_{\mu\pm1,\nu}^{\dagger}A_{\mu\pm1,\nu}+g_{\mu,\nu}^{\pm}A_{\mu\pm1,\nu}^{\dagger}P_{\mu,\nu}+ \text {H.C.}\bigg]\equiv \sum_{\mu,\nu} \hat{H}_{\mu,\nu}
\;,
\label{ham_cyl}
\end{equation}
where
$$
g_{\mu,\nu}^{\pm}=\pm {d}_{CV}\Phi(0) \frac {k_zS} {\sqrt {2 \pi}} \sin \left (k_z {L}/{2} \right )\sqrt{\frac{2\pi\hbar\omega (\mu,\nu, k_z)}{\epsilon \Omega (\mu,\nu, k_z)}}I_{\mu,\nu;\mu \pm 1,\nu} \;.
$$
If we diagonalize the Hamiltonian $\hat{H}_{\mu,\nu}$ using a $3\times 3$ Hopfield transformation, it will take the form
\begin{equation}
\hat{H}_{\mu,\nu}=\sum_{i=1}^3 E_i(\mu,\nu) \alpha_{\mu,\nu}^{(i)\dagger}\alpha_{\mu,\nu}^{(i)}\;,
\end{equation}
where $E_i$ $(i=1 -3)$ are the energies of the three polaritonic branches and the new operators $\alpha^{(i)}$ are linear combinations of the operators $A$ and $P$:
\begin{equation}
\alpha_{\mu,\nu}^{(i)}=\kappa_+^{(i)}(\mu,\nu)A_{\mu+1,\nu}+\kappa_-^{(i)}(\mu,\nu)A_{\mu-1,\nu}+\kappa_{ph}^{(i)}P_{\mu,\nu}\;,\qquad (i=1 -3)\;.
\label{alpha}
\end{equation}
%
%
\begin{figure}[!htb]\centering
   \begin{minipage}{0.48\textwidth}
     \includegraphics[width=.95\linewidth]{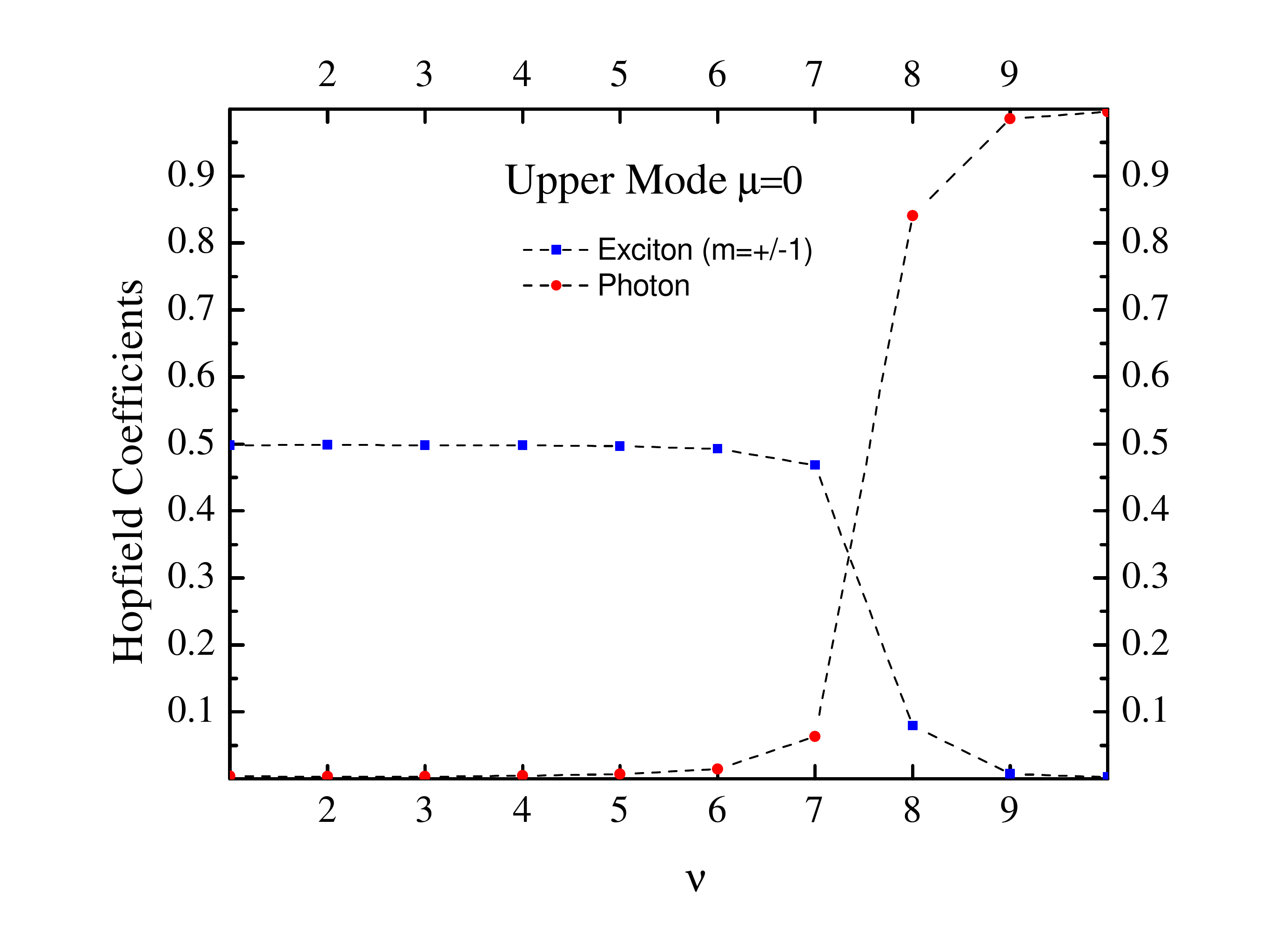}
   \includegraphics[width=.95\linewidth]{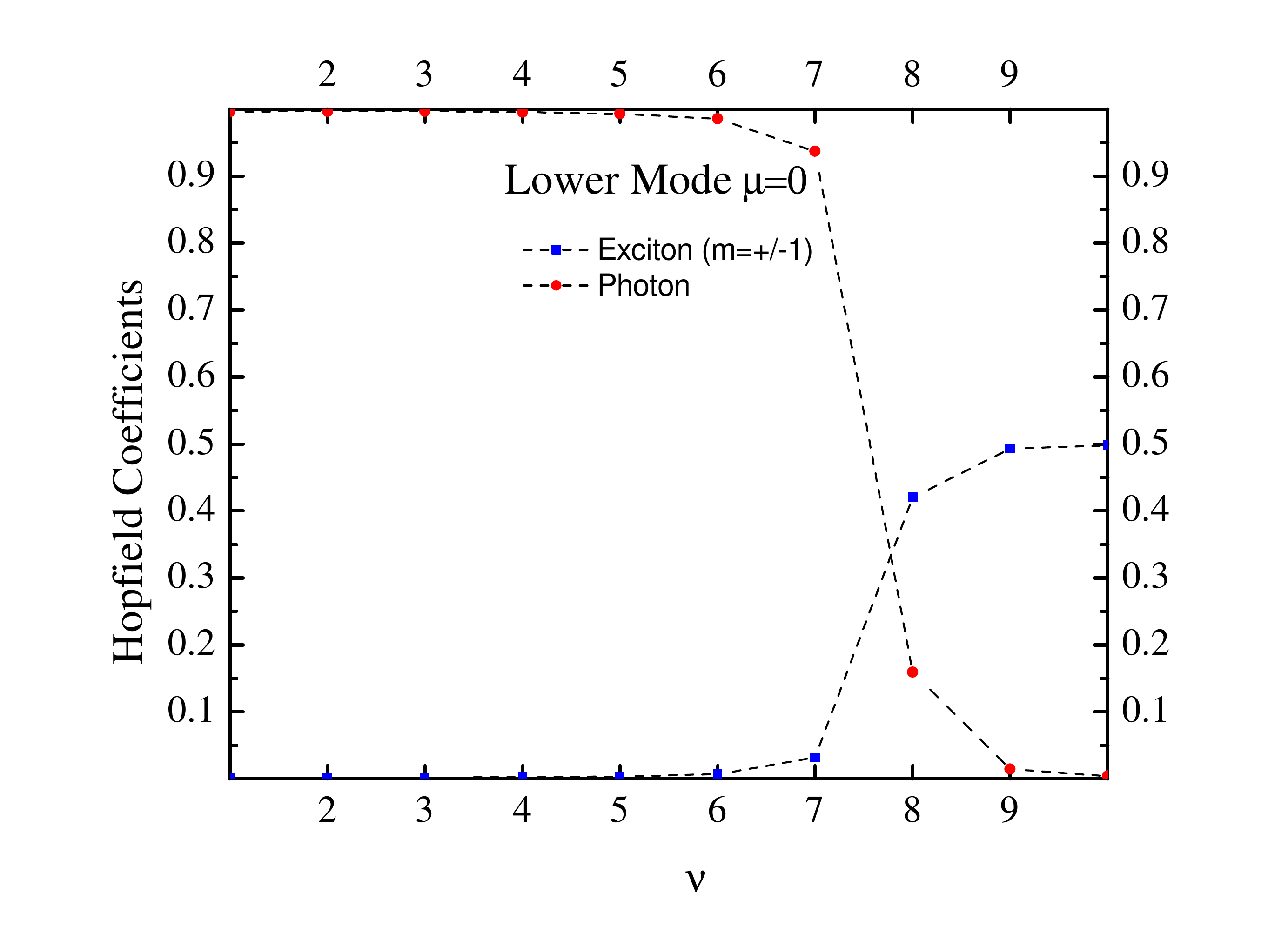}
   \end{minipage}
   \begin {minipage}{0.48\textwidth}
     \includegraphics[width=.95\linewidth]{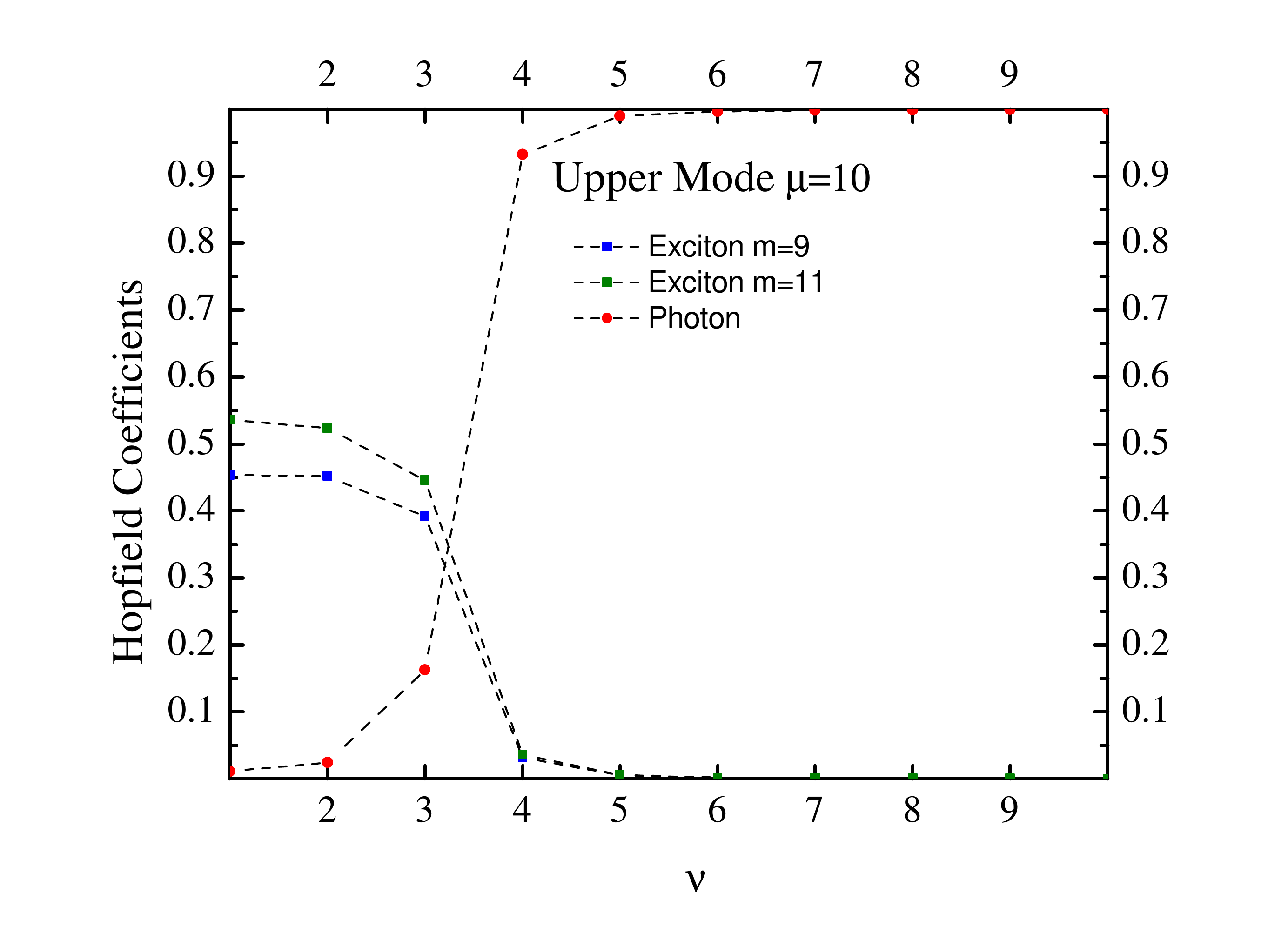}
   \includegraphics[width=.95\linewidth]{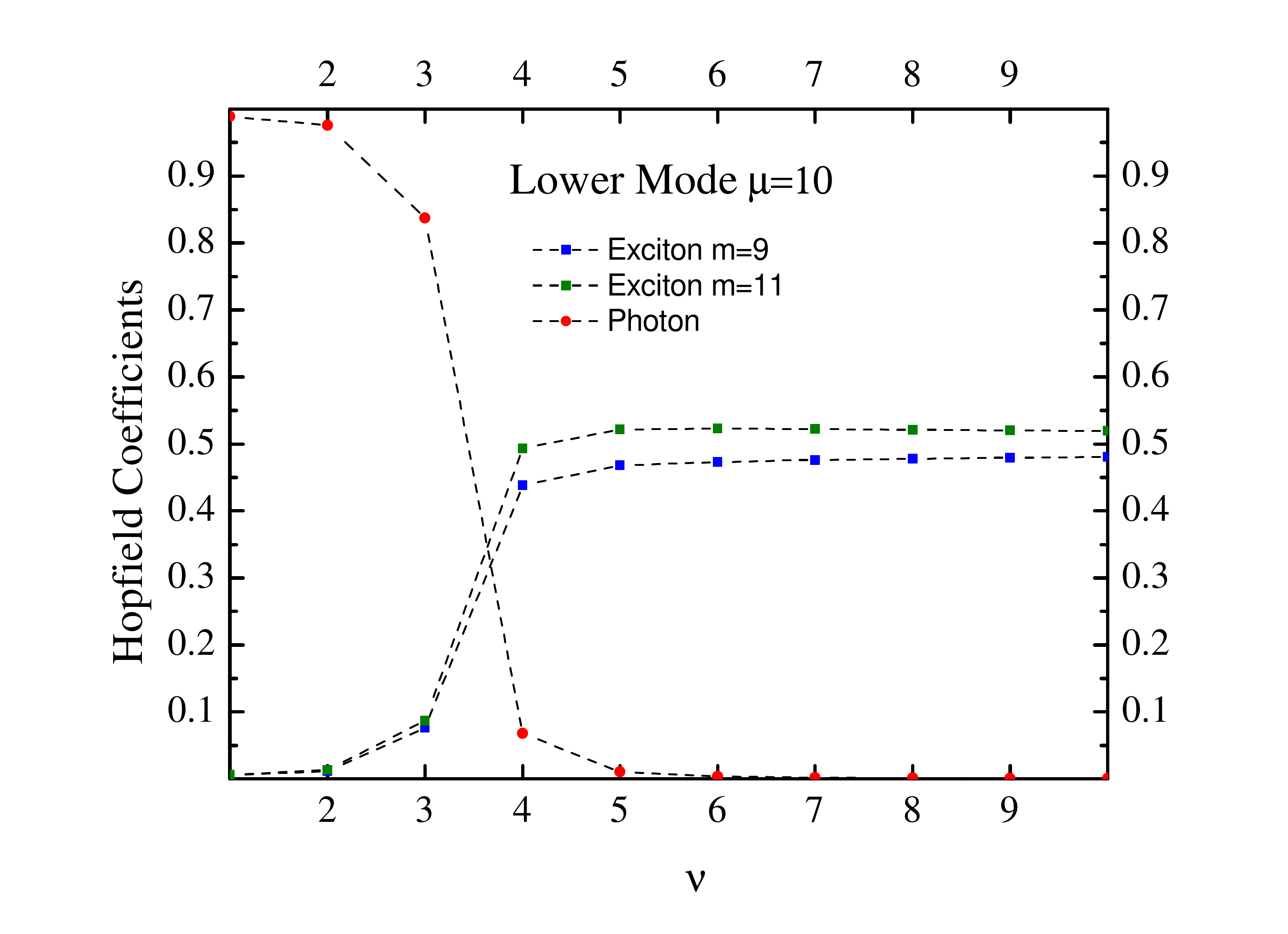}
      \end{minipage}
\caption{(Color online) Hopfield coefficients corresponding to the MC photon (red symbols and dash-lines) and two excitons ($\mu \pm 1$) (green and blue symbols and dash-lines) for the upper and lower polariton modes with $\mu=0$ and $10$. The parameters are the same as in Fig. \ref {Fig:classdispcurves}.}
\label{fig_e_and_kappa}
\end{figure}

In~(\ref {alpha}), $\kappa$ are the Hopfield coefficients, they constitute the $3\times 3$ Hopfield transformation matrix and are obtained by the diagonalization procedure of the Hamiltonian $\hat{H}_{\mu,\nu}$. The corresponding eigenenergies are
\begin{eqnarray}
\begin{aligned}
& E_1(\mu,\nu)=E_{ex} \; ;\\
&E_{2,3}(\mu,\nu)=\frac{E_{ex}+ \hbar \omega (\mu,\nu)}{2}\pm \sqrt{\frac{\big [E_{ex}-\hbar \omega (\mu,\nu)\big]^2}{4}+|g^-(\mu,\nu)|^2+|g^+(\mu,\nu)|^2} \; ,
\label{eigenvalues}
\end{aligned}
\end{eqnarray}
and the normalized Hopfield coefficients are (skipping the indices $\mu$ and $\nu$ for clarity):
\begin{eqnarray}
\begin{aligned}
&|\kappa_{ph}^{(1)}|^2=0 \; ,\qquad |\kappa_{\pm}^{(1)}|^2=\frac{|g^\mp |^2}{|g^+|^2+|g^+|^2} \; ;\\
&|\kappa_{ph}^{(2,3)}|^2=\frac{\big(E_{ex}-E_{2,3}\big)^2}{\big(E_{ex}-E_{2,3}\big)^2+|g^+|^2+|g^-|^2} \; ,\qquad
|\kappa_\pm^{(2,3)}|^2=\frac{|g^\pm|^2}{\big(E_{ex}-E_{2,3}\big)^2+|g^+|^2+|g^-|^2} \; .
\label{Hkappas}
\end{aligned}
\end{eqnarray}

As seen from Eqs.~(\ref {eigenvalues}) and  (\ref {Hkappas}), the first mode is purely excitonic, while $E_{2,3}(\mu,\nu)$ correspond to two polariton branches.  As we can see from Fig.~\ref {Fig:separations}, the polariton eigenmode energies obtained within the reduced quantum model reproduce qualitatively quite well those calculated  within the classical picture including all exciton-photon interactions, even thogh the former underestimates the separation between the upper and lower polariton branches. The minimum separation, i.e. the Rabi splitting is approximately equal to 100 mEv and almost independent of $\mu$  (see Fig.~\ref {Fig:separations}). This is somewhat smaller than the value calculated for a TMD layer with the same parameters inserted in a planar MC~\cite{Vasilevskiy2015} but closer to the experimentally measured values.~\cite{Ye2015}  

The dependence of the Hopfield coefficients on $\nu$ is shown in Fig.~\ref {fig_e_and_kappa} for $\mu =0$ and 10. For $\mu=0$ the two Hopfield coefficients measuring the fraction of excitons with $m=\pm 1$ are equal, while for $\mu =10$ the contributions of $\mu =9$ and $\mu =11$ are slightly different below the crossing point of the dispersion lines of the uncoupled photons and excitons.

\section{Polariton Densities of States}

\subsection{Total and Projected Densities of States}
\begin{figure}[H]
\includegraphics[width=.5\linewidth]{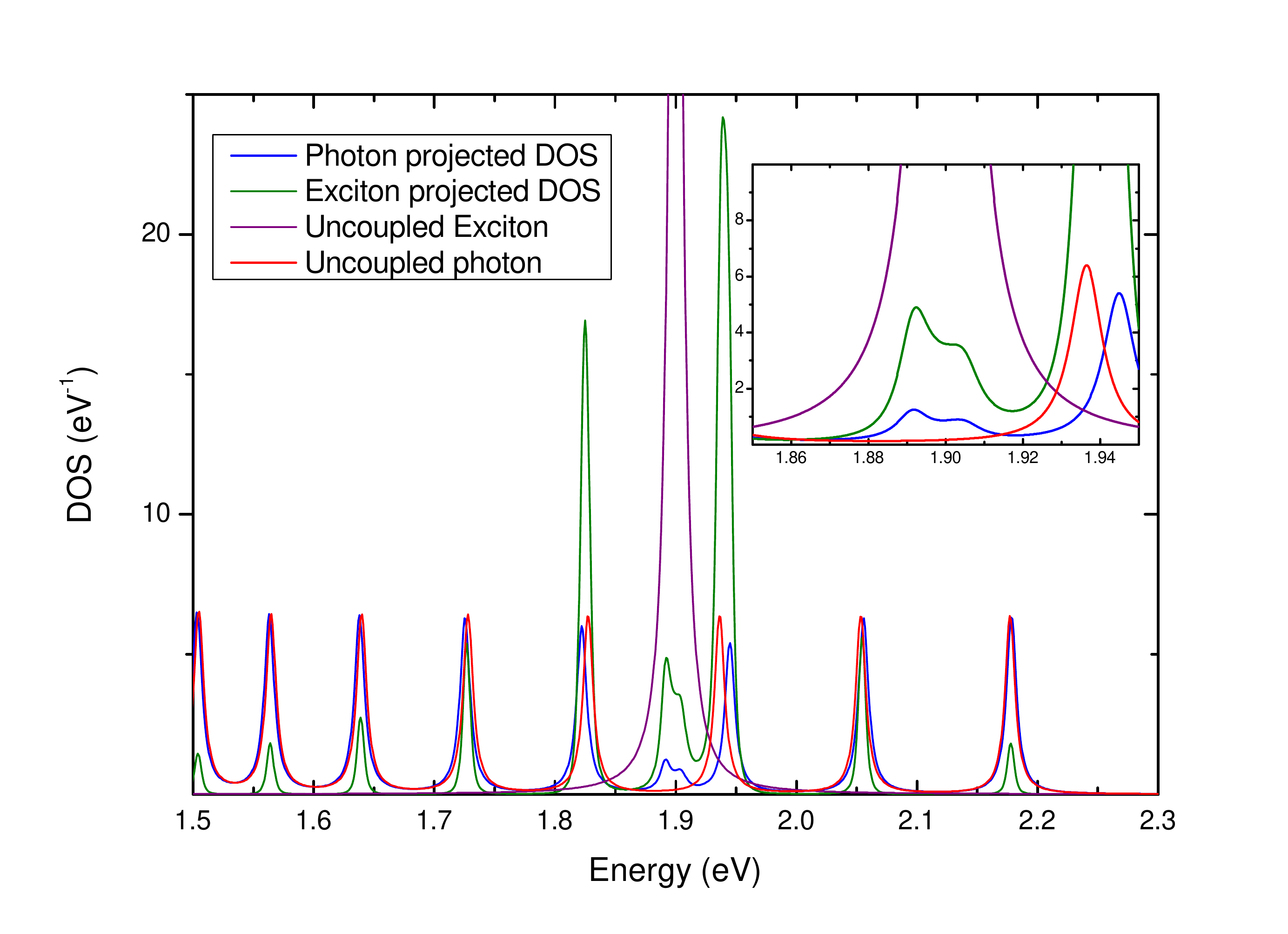}
\centering
\caption{ (Color online) Dependence of the DOS,  $\rho ^{(\mu)}$, for empty cavity (defined similar to Eq. (\ref{DOS}) but without sum over $i$, red curve) and DOS for cavity with TMD layer projected over photonic ($\rho_{ph}^{(\mu)}$, Eq. (\ref{PDOS}), blue curve)  and excitonic (green curve) subspaces for $\mu=0$. Also shown is the uncoupled exciton peak. 
All densities of states are shown normalized to unity, i.e. divided by $\int {\rho^{(\mu)}(E)dE}$. The inset shows a zoom into the energy interval near the avoided crossing.
\label{fig_DOS}
}
\end{figure}

The angular momentum, $\mu$, still is a well-defined quantum number for polariton modes and, owing to our approximation, so is the radial number $\nu$.
The density of polariton states (DOS) with a certain angular momentum $\mu$ (and for the lowest $k_z$) can be defined as
\begin {equation}
\rho ^{(\mu)}(E)=\sum_{\nu,\,i}\delta(E_i(\mu,\nu)-E)\;,
\label{DOS}
\end{equation}
where $i$ stands for the three polaritonic branches. The same definition (without sum over $i$) can be used for purely photonic modes in empty cavity. 
It is interesting to compare the latter to the polariton DOS projected onto the photonic subspace, which is calculated by weighting each polariton mode by the corresponding (photon) Hopfield coefficient:
\begin{equation}
\rho_{ph}^{(\mu)}(E)=\sum_{\nu,\,i}\vert \kappa_{ph}^{(i)}(\mu,\nu)\vert ^2\delta(E_i(\mu,\nu)-E)\;.
\label{PDOS}
\end{equation}
The projected density of states (PDOS),  $\rho_{\pm }^{(\mu)}(E)$, for excitons with angular momenta $\mu \pm 1$, respectively,
is defined similar to (\ref{PDOS}). Note that the sum of the three PDOS functions gives the total density of states, i.e. $\rho^{(\mu)}=\rho_{ph}^{(\mu)}+\rho_{+}^{(\mu)}+\rho_{-}^{(\mu)}$. 
The dependencies of the functions $\rho_{ph}^{(\mu)}$, $\rho_{\pm}^{(\mu)}$ (coinciding for $\mu =0$) on $E$ are displayed in Figs.~\ref{fig_DOS} along with the DOS for uncoupled cavity photons and excitons.
%
%

From Fig.~\ref{fig_DOS} it can be seen that  the function $\rho_{ph}^{(0)}$ follows the trends of $\rho ^{(0)}$ for pure photons in a broad range of energy values below the crossing point ($E<$1.8 eV). Close to the crossing point (1.9 eV), the polaritons are an almost equal-fraction admixture of MC photons and two exciton species with angular momenta $\mu \pm 1$. It results in a non-zero photonic PDOS within the "gap" of photon DOS with $\mu=0$.
For $E>2.0$ eV,  the upper polariton branch is almost photon (see Fig.~\ref {fig_e_and_kappa}), therefore, $\rho_{ph}^{(0)}$ and the empty cavity DOS again are similar.

\subsection{Local Density of States}

The local density of states (LDOS), i.e. space resolved DOS of photons is important since it determines the variation of the spontaneous decay rate of a point emitter placed in the cavity; it is defined by weighting each photon mode with squared local magnitude of the electric field.~\cite{Hecht-Novotny} Here we deal with mixed photon-exciton modes, therefore, it makes sense to consider the local density of states {\it projected onto the photon subspace} defined as follows:
\begin{equation}
\rho_{loc}^{(\mu)}(E;r,z)=\sum_{\nu,\,i}\vert\vec{\mathbb{E}}(\mu,\nu;r,\phi,z)\vert^2\vert \kappa_{ph}^{(i)}(\mu,\nu)\vert ^2\delta(E_i(\mu,\nu)-E)\;.
\label{LDOS}
\end{equation}
The electric field amplitude in~(\ref{LDOS}) corresponds to one photon and is expressed through the mode volume according to Eq.~(\ref{E0}).
Employing Eq.~(\ref{energycons}) it follows that the integration of $\rho_{loc}^{(\mu)}(E;r,\phi,z)$ over the MC volume yields the total energy of the photonic subsystem with angular momentum $\mu$, therefore, we have
\begin{equation}
\frac \epsilon {2\pi E}\int \rho_{loc}^{(\mu)}(E;r,z)d^3r=\rho_{ph}^{(\mu)}(E)\;.
\label{normalization}
\end{equation}

\begin{figure}[!htb]
\centering {
   \begin{minipage}{0.5\textwidth}
    \includegraphics[width=1\linewidth]{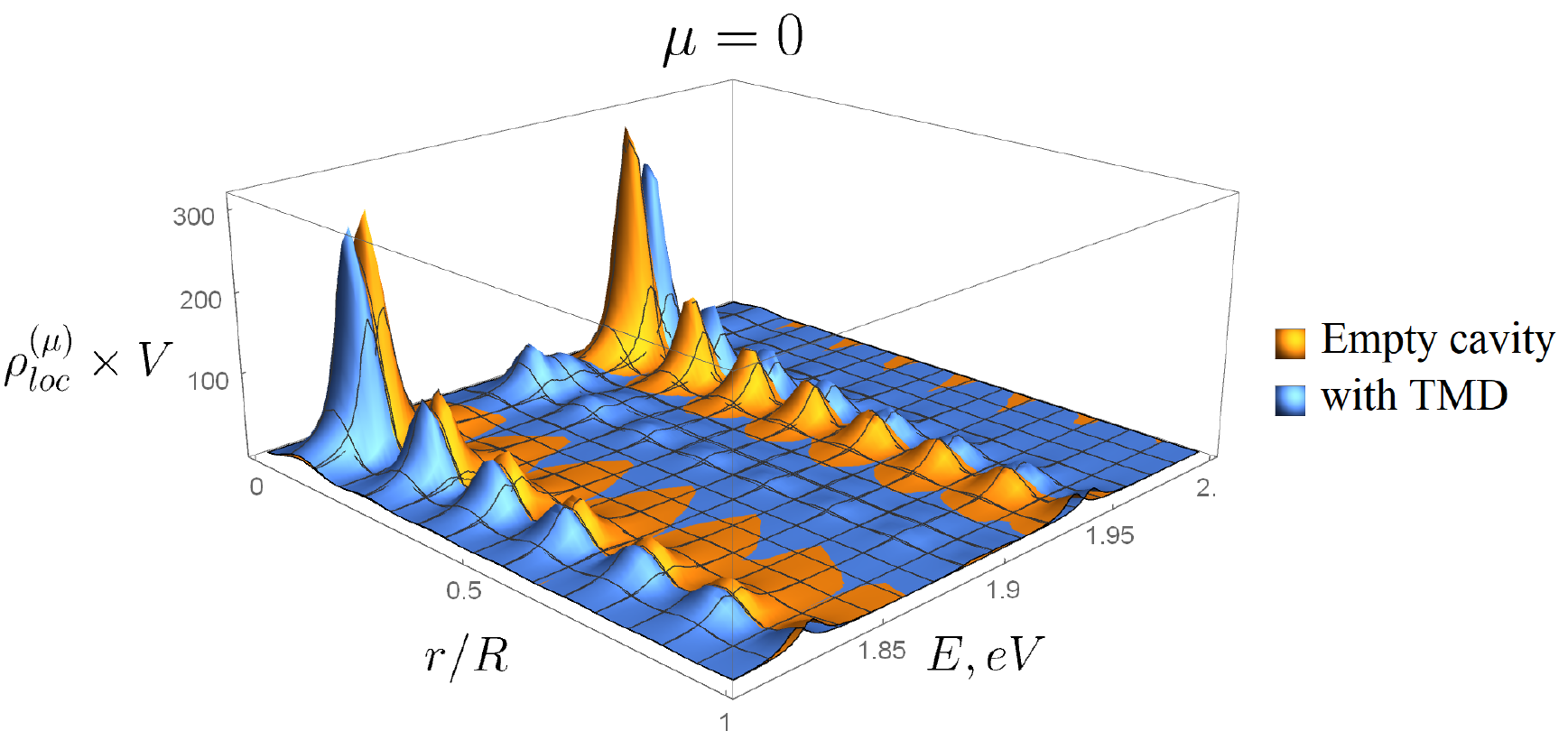}
   \end{minipage}
   \begin {minipage}{0.4\textwidth}
    \includegraphics[width=1\linewidth]{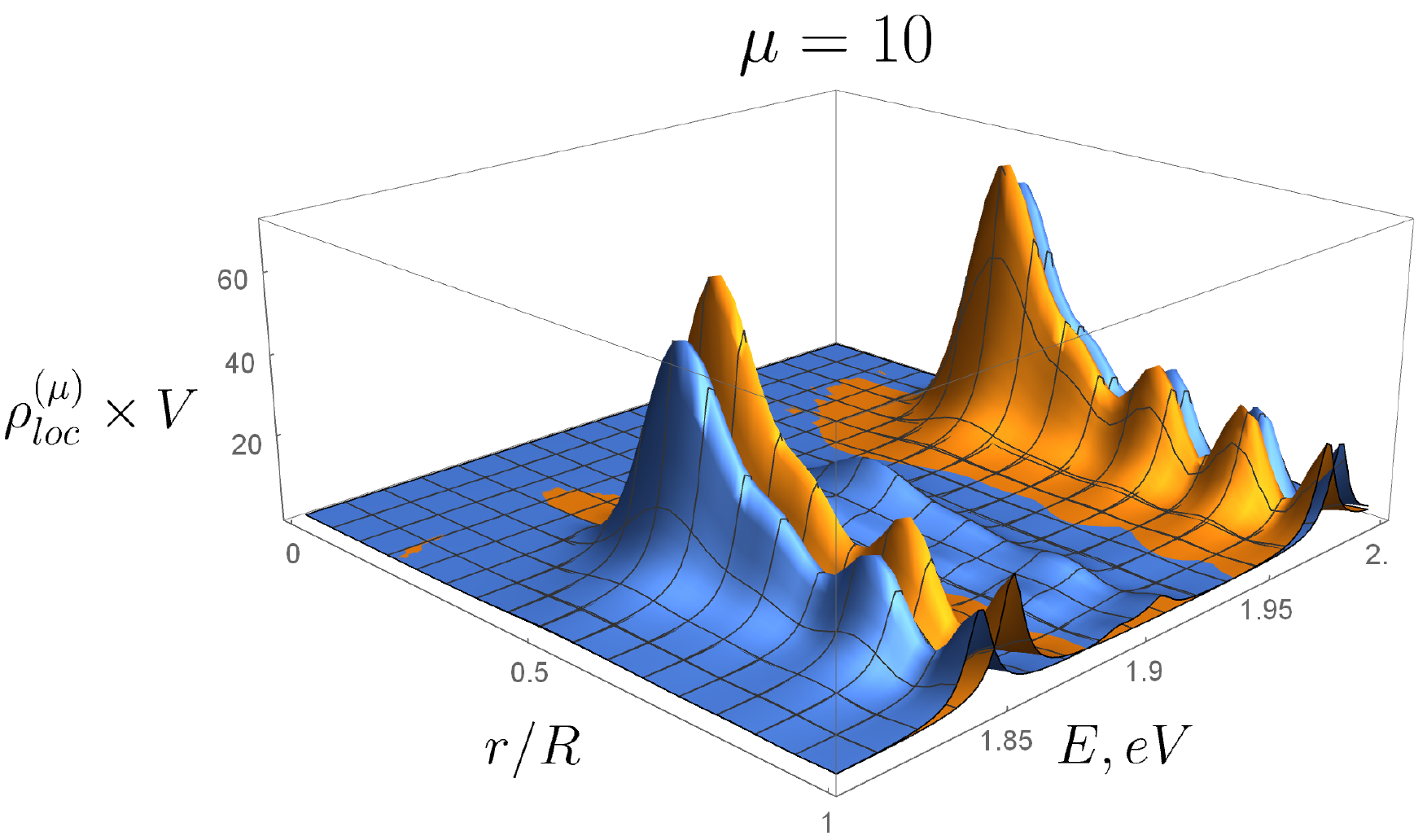}
      \end{minipage}
\includegraphics[width=0.95\linewidth]{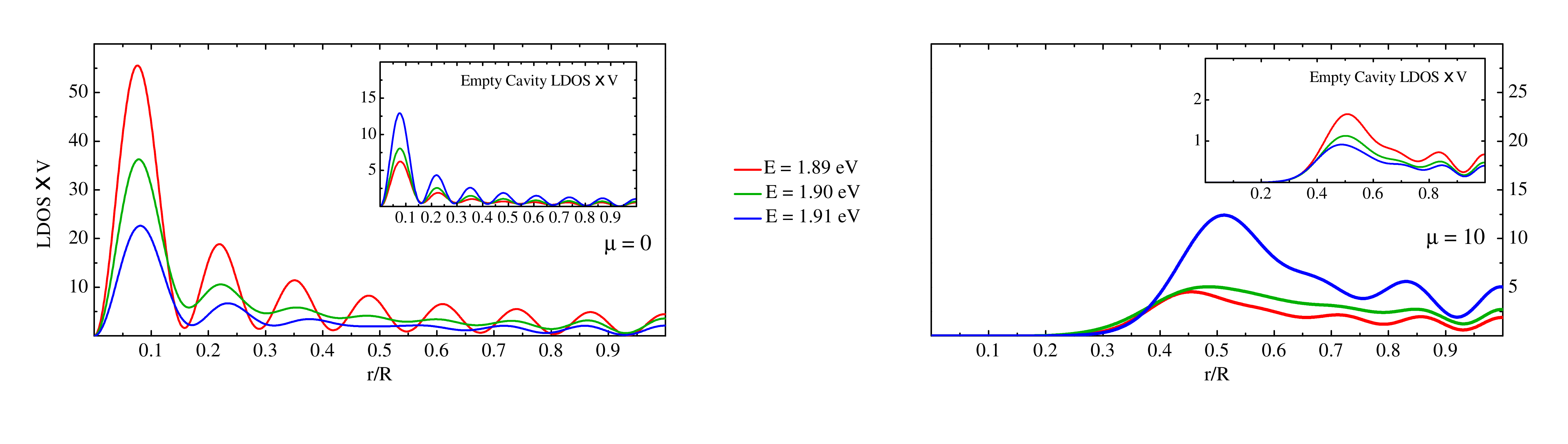}
}
\caption{(Color online) Dependence of the local density of states, $\rho_{loc}^{(\mu)}(E;\,r,\,z=0)\times V$ (where $V=\pi R^2L$), on the normalized radius $r/R$, for empty cavity and for cavity with TMD [where it is the polariton photon-projected LDOS, Eq. (\ref {LDOS})], for $\mu=0$ (left) and $10$  (right). The lower panels highlight the  redidstribution of LDOS due to the TMD layer for the energies near the avoided crossing point. The parameters are the same as in Fig. \ref {Fig:classdispcurves}.}
\label{fig_Photon-proj LDOS}
\end{figure}

For the implementation that we have in mind, discussed below, it is particularly interesting to calculate LDOS at the TMD layer, i.e. for $z=0$. In Fig.~\ref{fig_Photon-proj LDOS} it is shown the dependence of the LDOS, $\pi R^2 \rho_{loc}^{(\mu)}(E;r,z=0)$, upon $r$ and energy in the vicinity of the crossing point. As it can be seen from the plots, the local density of states depends strongly on both the energy and the radial position and it is redistributed owing to the presence of the TMD layer. 
%
%

\section{Point Emitter Attached to the TMD Layer: the Purcell effect}
In the polariton framework, the exciton does not decay radiatively, instead, the strongly coupled system oscillates between the state with and without exciton and the energy is reversibly transferred from the exciton to photon and {\it vice versa}, the effect usually referred to as vacuum Rabi oscillations.~\cite{Andreani2014} Yet, in a real cavity some photons escape (and, accordingly, some radiative decay of excitons does occur) giving rise to {\it polariton photoluminescence}, which can been observed experimentally. However, studies of MC-embedded semiconductor quantum wells (QWs) showed that this emission mechanism is inefficient and the luminescence is dominated by excitons bound to impurities or defects, with the QW emission line redshifted with respect to the polariton modes.~\cite{Andreani2014}  

Here we shall consider a point emitter (PE) located within the TMD layer or very close to it, however, without any direct interaction with the TMD excitons.
We can think of a trapped exciton in TMD~\cite {Tonndorf2015}  or a point defect  in an  h-BN layer~\cite{Tran2016} attached to the TMD layer or even a nanocrystal quantum dot (QD) placed on top of it.~\cite{Raja2016,Prins2014}
We shall assume that it yields an optical transition at  energy, $ \hbar \omega _0$, sufficiently close but not coinciding with $E_{ex}$.
The emitter interacts with the MC photon modes, which are affected by the strong coupling to the 2D excitons and it should influence emitter's properties.
We shall assume that the PE is in the so called weak coupling regime.~\cite {Kavokin_MCs,Sargent-III} 

The spontaneous emission rate of a point emitter located within a homogeneous infinite dielectric with refractive index $\eta =\sqrt {\epsilon}$ is~\cite {Hecht-Novotny,Sargent-III} 
\begin{equation}
 \Gamma _0 =\frac {4 \omega _0^3 d_0^2}{3\hbar c^3}\eta
\label {Gamma0}
\end{equation}
where $d_0$ is the dipole moment matrix element of the PE optical transition and
and $\omega_0$ is the emission frequency.
In the weak coupling regime, the spontaneous emission rate placed in an ideal photonic microcavity is enhanced or inhibited depending on the photonic LDOS value in the position of the emitter,~\cite {Hecht-Novotny}
\begin{equation}
 \Gamma (\vec{r}) =\frac {4 \pi ^2 \omega _0 d_0^2}{\hbar }\rho ^{||}_{loc} (\hbar \omega _0, \vec{r})\;,
\label {Gamma}
\end{equation}
where
\begin{equation}
\rho ^{||}_{loc}(E;\vec{r})=\sum_{\mu,\,\nu}\vert\vec{{u}}(\mu,\nu;r,\phi,z)\cdot \vec n_d\vert^2\delta(\hbar \omega(\mu,\nu)-E)
\label{LPDOS}
\end{equation}
is the {\it local projected} density of photon states (LPDOS). 
Here $\vec {n_d}$ is the unit vector along the emitter's dipole moment, $\vec {d_0}$, $\omega(\mu,\nu)$ is the frequency of the cavity mode with certain $\mu $ and $\nu $ (for simplicity we consider only modes with the lowest $k_z$), and
$$
\vec{{u}}(\mu,\nu;r,\phi,z)=\frac \eta {\sqrt {2\pi \hbar \omega(\mu,\nu)}}\vec{\mathbb{E}}(\mu,\nu;r,\phi,z)\
$$
(notice that it is normalized to unity). Note that we removed the factor $1/3$ in Eq. (\ref {Gamma}) and, correspondingly, the factor of $3$ in the LDOS definition in Eq. (\ref {LDOS}), compared to Ref.~\onlinecite {Hecht-Novotny}. Taking average over possible orientations of the dipole moment yields:
$$
\left\langle \vert\vec{{u}}(\mu,\nu;r,\phi,z)\cdot \vec n_d\vert^2\right\rangle
=\frac {\epsilon p^{-1}(\mu,\nu;r,\phi,z)}{{2\pi \hbar \omega(\mu,\nu)}}\vert\vec{\mathbb{E}}(\mu,\nu;r,\phi,z)\vert^2
$$
where $p$ is a numerical factor (e.g. $p=1$ if the dipole moment is aligined with the considered mode and  $p=3$ if its orientation is isotropic in 3D space). Let $\bar p^{-1} =\langle p^{-1}(\mu,\nu;r,\phi,z) \rangle$ be its average value over all its arguments, a number of the order of unity. Then we can express the local emission rate through the usual local density of states (LDOS) as follows: 
\begin{equation}
 \Gamma (\vec{r}) =\Gamma _0 \frac {3\epsilon \bar p^{-1} \rho ^{||}_{loc} (\hbar \omega _0, \vec{r})}{2\pi \hbar \omega _0 \rho _0(\hbar \omega _0)}
\label {Gamma-2}
\end{equation}
where $ \rho _0(\hbar \omega _0)=\omega _0^2\eta ^3/(\pi ^2 \hbar c^3)$ is the photon DOS in an infinite homogeneous dielectric medium.
Equation (\ref {Gamma-2}) expresses the so called the Purcell effect,~\cite{Purcell1946} which can be controlled by manipulating the photonic LDOS in nanostructures.~\cite {Noda2007,Pelton2015}
Nowadays it finds applications in nano-optical spectroscopy, nanolasers, or broadband single-photon sources.~\cite {Sauvan2013} 

If the emitter is located at $z=0$, the relevant LDOS is written as follows:
\begin{equation}
\rho ^{||}_{loc}(\hbar \omega _0;r)=\frac {2\pi \hbar \omega _0}{\epsilon }\sum_{\mu,\,\nu}\Omega ^{-1}(\mu,\nu)
\left (\frac {k _z}{q_{\mu \nu}}\right )^2
\left \{\left [J^\prime _\mu(q_{\mu \nu}r)\right ]^2 + \left [\mu J _\mu(q_{\mu \nu}r)/(q_{\mu \nu}r)\right ]^2 \right \}\delta(\hbar \omega(\mu,\nu)-\hbar \omega _0)\;,
\label{LDOS-2}
\end{equation}
Taking into account losses due to imperfect mirrors (with the loss rate $\gamma _{loss}$), the $\delta $-function in  (\ref {LDOS-2} is replaced by a Lorentzian function according to~\cite {Sargent-III}
\begin{equation}
\delta(\hbar \omega(\mu,\nu)-\hbar \omega _0)\rightarrow \frac 2\pi \frac 1{\hbar \omega _0} 
\frac {Q}{1+4Q^2\left [ \omega (\mu,\nu)/\omega _0 -1\right]^2}
\;,
\label{lorentzian}
\end{equation}
where $Q=\gamma _{loss} / \omega _0$ and the $\delta $-function is recovered in the limit $Q\rightarrow \infty$.

Noting that 
$$
2\pi \hbar \omega _0 \rho _0(\hbar \omega _0)=16 \pi ^2 \left (\frac \eta {\lambda _0} \right )^3
$$
where $\lambda _0$ is the emission wavelength in vacuum, we can rewrite Eq. (\ref {Gamma-2}) as follows:
\begin{equation}
\frac { \Gamma ({r}) }{\Gamma _0}=\sum _{\bar \mu, \bar \nu}\left [\frac {3Q} {4\pi ^2}\Omega ^{-1}(\bar \mu ,\bar \nu)\left (\frac {\lambda _0} \eta \right )^3 \right ] \bar p^{-1} \varphi (\bar \mu ,\bar \nu;r)\;,
\label {Gamma-3}
\end{equation}
where we have assumed exact resonance between the emission frequency and a cavity mode $\bar \mu, \bar \nu$, so that the sum is over all modes obeying the relation $\omega (\bar \mu, \bar \nu) =\omega _0$, and 
$$
\varphi (\bar \mu,\bar \nu;r)=\left (\frac {k _z}{q_{\bar \mu \bar \nu}}\right )^2\left \{\left [J^\prime _{\bar \mu}(q_{\bar \mu \bar \nu}r)\right ]^2 + \left [\bar \mu J _{\bar \mu}(q_{\bar \mu \bar \nu}r)/(q_{\bar \mu \bar  \nu}r)\right ]^2 \right \}\;.
$$
Therefore, the emission rate is proportional to the quality factor, $Q$, and inveresely proportional to the (resonance) mode volume, while its spatial dependence is determined by the function $\varphi (\bar \mu,\bar \nu;r)$. The factor $\bar p^{-1}$ takes into acount the emission polarization properties of the emitter. 
The term in square brackets in Eq. (\ref {Gamma-3}) is called the Purcell factor (originally introduced by Purcell~\cite{Purcell1946} with the cavity volume, $V$, in place of the mode volume, $\Omega$) and it measures the maximum spontaneous emission acceleration that can be achieved by placing the emitter in a cavity, in the weak coupling regime. The Purcell factor has been generalized to take into account the effects such as photon confienement, absorption and MC material's dispersion,~\cite{Sauvan2013} so Eq. (\ref {Gamma-3}) can be seen as a generalization taking into account the cylindrical geometry. 

\begin{figure}[!htb]\centering
   \begin{minipage}{0.55\textwidth}
    \includegraphics[width=.95\linewidth]{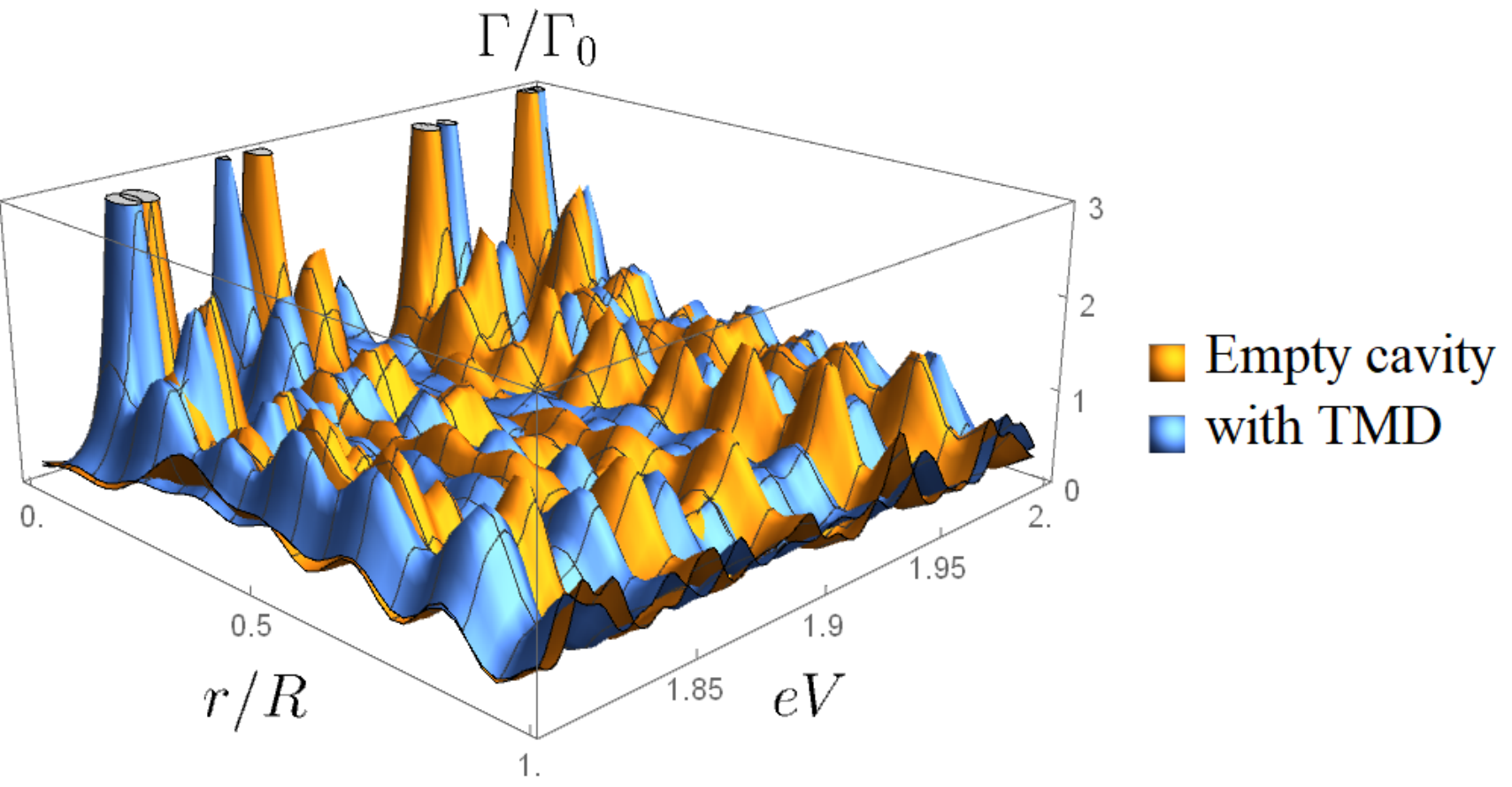}
   \end{minipage}
   \begin {minipage}{0.41\textwidth}
    \includegraphics[width=.95\linewidth]{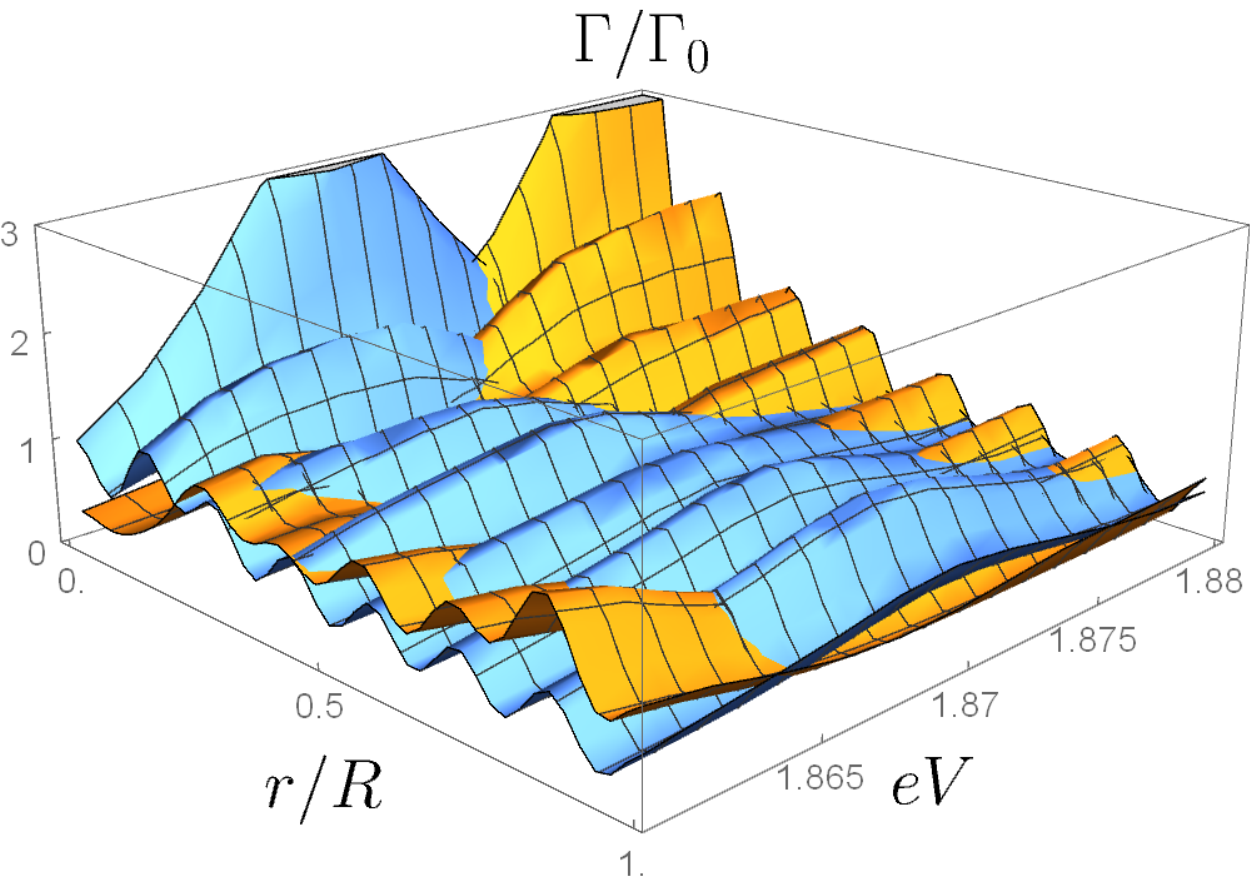}
      \end{minipage}
 \caption{(Color online) Energy and position dependence of the local emission rate of a point emitter located in the $z=0$ plane calculated for empty cavity and for cavity with TMD. The right panel highlights the region near 1.87 eV where the Purcell effect enhancement due to the TMD layer is seen. We assumed $\bar p^{-1} =1$ and other parameters are the same as in Fig. \ref {Fig:classdispcurves}.}
\label{fig:Purcell}
\end{figure}

For a cavity with embedded TMD layer, the local emission rate is expressed through the local polaritonic density of states projected onto the photon subspace, defined in the previous section [Eq. \eqref {LDOS}] as follows: 
\begin{equation}
 \Gamma (\vec{r}) =\Gamma _0 \frac {3\epsilon \bar p^{-1} \sum _\mu {\rho ^{(\mu )}_{loc} (\hbar \omega _0, \vec{r})}}{2\pi \hbar \omega _0 \rho _0(\hbar \omega _0)}\;.
\label {Gamma-4}
\end{equation}
Figure \ref{fig:Purcell} illustrates how the local emission rates changes with energy and emitter's position within the $z=0$ plane inside either an empty cavity or a cavity with inserted TMD layer.

%
%
From this figure we can see that local emission rate is redistributed both in energy and in space due to the presence of the TMD layer in the cavity.
Such an engineering of $\Gamma (\vec{r})$ can be useful for stimulating a point emitter tuned to a particular wavelength while pumping it through a cavity mode. If it is a trapped exciton in the TMD layer, the red shift of the photon-projected polariton LDOS with respect to the bare 2D exciton energy (1.9 eV in our example) is desirable indeed. Excitation can be performed through the upper polariton branch lying above 1.9 eV (see Fig. \ref {Fig:classdispcurves}). The excitonic part of such a polariton can be trapped by a defect or impurity. Angular momentum is not conserved for trapped excitons and their emission, after being excited, can be proportional to the local photonic DOS irrespective of $\mu$. Yet, as Fig. \ref {fig_Photon-proj LDOS} shows, LDOS, for the same energy, peaks at different distances from cavity's centre and it will enhance the emission of photons with certain angular momenta depending on the position of the emitter.

\section{Conclusion}
Most of this work has been dedicated to the properties of exciton-polaritons that arise due to strong coupling of cylindrical cavity photon modes to 2D excitons confined in a TMD semiconductor layer placed inside the cavity.
Such a system is feasible and may be used in on-chip optoelectronics,~\cite {Ye2015} and has advantages in comparison with planar MCs because of the lateral confinement of the EM field permitting to decrease the mode volume and therefore enhance light matter interaction.
This exciton-mediated coupling is intrinsically strong enough for single- and few-layer TMDs and its enhancement by taking advantage of the MC effect can make these structure competitive with the traditional semiconductor materials used in optoelectronics.~\cite{Liu2015,Flatten2016,Lundt2016}
The cylindrical geometry allows, in principle, for separate excitation of the cavity modes with different angular momenta ($\mu $), which have distinct radial distribution of the EM field. For large $\mu$, so called whispering gallery modes can be formed with the field amplitude peaking near
the lateral walls; such modes have been observed in II-VI semiconductor micropillars.~\cite{Sun2008,Jakubczyk2012} The eigenfrequencies of polariton modes with different angular momenta and radial number $\nu $ have been obtained and presented in the form of dispersion curves and DOS in Sections III and IV respectively.
Also, we showed that the simplified quantum model of the polariton modes taking into account coupling of a photon cavity mode with certain $\mu $ and $\nu $ only to two excitons with the (centre of mass) quantum numbers $m=\mu \pm 1$ and the $n=\nu$, providing a reasonable description of them and yielding an explicit form of the Hopfield coefficients.

The coupling with the 2D layer causes a strong enhancement of the photonic density of states in the cavity, as we have demonstrated by calculating the photon-projected
(i.e. weighted by the corresponding Hopfield coefficient) DOS for certain energies. In particular, it peaks near the exciton resonance energy (see  Fig. \ref {fig_Photon-proj LDOS}) that, for a micrometer-diameter cavity is practically independent of $m$ and $n$.
The local photon-projected DOS shows a similar behaviour as a function of energy and it testifies the $\mu$-dependent spatial distribution of the photonic field outlined above.
This quantity determines the spontaneous emission rate of a quantum emitter placed in the cavity, so the photoluminescence intensity of such an emitter will depend on its position or, seen from another point of view, on the direction of the light beam exciting the microcavity (by selecting cavity polariton modes with different $\mu $). Such effects have been observed for micropillar cavities made of a bulk semiconductor, using either point defects~\cite{Sun2008} or QDs~\cite{Jakubczyk2012} as emitters. Fot 2D materials, such an emitter can be a trapped exciton in the TMD layer (emission energy below the free exciton one, $E_{ex}$ or in a nearby h-BN layer~\cite{h-BN_emission_2016} (emission above $E_{ex}$, in which case an upconversion process~\cite{Santos2008,Rakovich2009} is required for the excitation). We hope that the presented calculated results will help devising such experiments for microcavities with TMDs, in particular, aimed at building controllable single photon emitters. Single photon emission has been observed for various TMDs~\cite {Tonndorf2015} and also from h-BN~\cite{h-BN_emission_2016}, presumably originated  from the radiative recombination of excitons trapped on material defects.~\cite {Zheng2018,h-BN_emission_2016} The location of such an emitter is random and, depending on the position it will couple to polariton modes with different angular momenta. This feature may be used to excite the emitters space-selectively. On the other hand, emitters with a relatively large absorption cross-section (such as QDs~\cite{Rakovich2009}) placed in the system in a controlled way may be used to criate exciton-polaritons as suggested in Refs.~\onlinecite{Cartstein2015,Prins2014} or, in the case of cylindrical cavity,  a particular class of them with certain angular momenta.

\section*{Acknowledgments}
We are grateful to Dr. Fernando de Le\'on for helpful discussion.
Funding from the European Commission, within the project "Graphene-Driven Revolutions in ICT and Beyond" (Ref. No. 696656), and from the Portuguese Foundation for Science and Technology (FCT) in the framework of the PTDC/NAN-OPT/29265/2017 “Towards high speed optical devices by exploiting the unique electronic properties of engineered 2D materials” project
the Strategic Funding UID/FIS/04650/2019 is gratefully acknowledged. C. T-G acknowledges support
from the Brazilian Agency CNPq.

\appendix

\section{\label{appendix A} Comparison of field profiles: ideal cavity {\it vs} exact solution for infinite cylinder}
%
%
\begin{figure}[h]
\includegraphics[width=.45\linewidth]{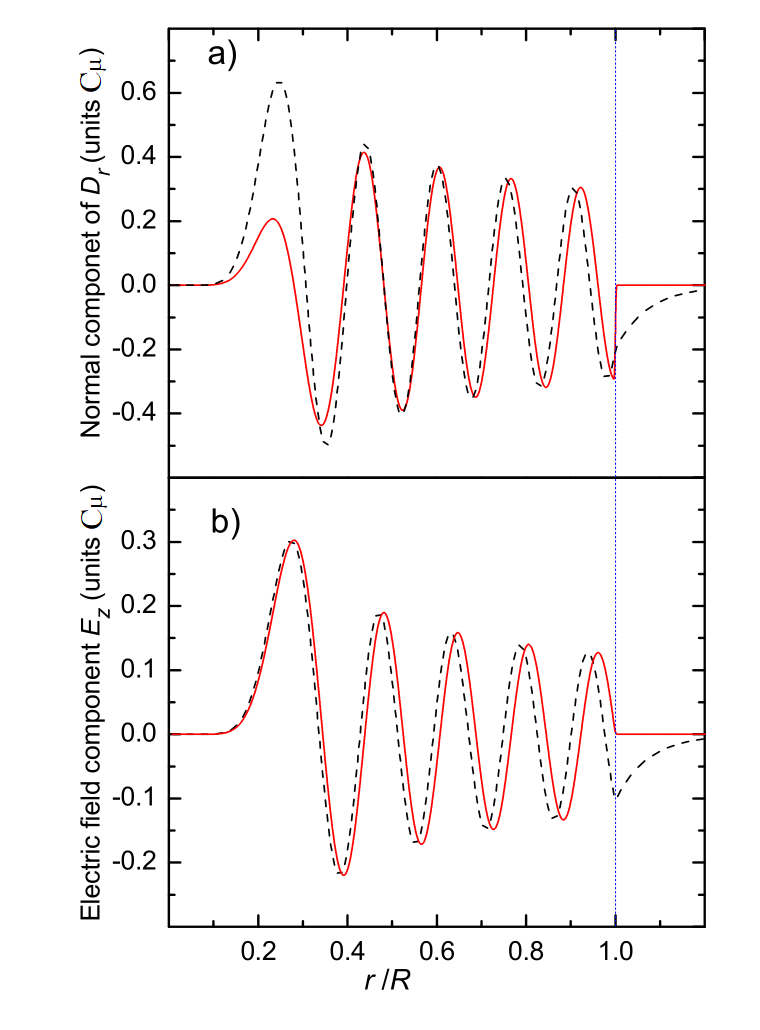}
\caption{Comparison of the field components' profiles $E_z$ and $D_r$ from exact solution, Eq.~(\ref{EcuSecular}) (dashed lines) and from the ideal cavity model, Eq.~(\ref{34}) (full red lines) for $c k_z=$ 4 eV.  The radius of the cylinder $R=3\;\mu$m, $\epsilon_1=$1, $\epsilon_2=$3.4, and the height $L=0.3\;\mu$m.}
\label{Fig:profiles}
\end{figure}
The EM fields used in this work correspond to an ideal cavity with its boundaries being perfect mirrors. However, in reality the contrast of the dielectric constant at the boundaries is finite.
We shall compare these field profiles obtained within the ideal cavity model to exact solution of Maxwell's equations in the case of infinite cylinder where analytical solution is possible considering finite dielectric constant (and therefore non-zero field amplitudes) outside of the cylinder.

The equations that relate the transverse to the longitudinal components of the vector fields are
\begin{equation}
-\frac{i\omega\epsilon}{c}{\vec{\mathbb{E}}_T}=
\left(
\begin{array}{c}
\frac{1}{r}\frac{\partial H_z}{\partial \phi}-\frac{\partial H_{\phi}}{\partial z}\\
\frac{\partial H_r}{\partial z}-\frac{\partial H_z}{\partial r}
\end{array}
\right)\;;\;\;\;\;\; \;\;\frac{i\omega}{c}{\vec{\mathbb{H}}_T}=\left(
\begin{array}{c}
\frac{1}{r}\frac{\partial E_z}{\partial \phi}-\frac{\partial E_{\phi}}{\partial z}) \\
\frac{\partial E_r}{\partial z}-\frac{\partial E_z}{\partial r}
\end{array}
\right)
 \;.
\label{amplitude}
\end{equation}
Due to the translational and axial symmetry of the cylindrical microcavity in Fig.~\ref{scheme}, the electromagnetic fields $\vec{\mathbb{E}}$ and $\vec{\mathbb{H}}$  can be cast as
\begin{equation}
\vec{\mathbb{E}}(r,\phi,z)=\vec{E}(r)\Psi(\phi,z)\;;\;\;\;\;
\vec{\mathbb{H}}(r,\phi,z)=\vec{H}(r)\Psi(\phi,z)\;,
\label{fiedl}
\end{equation}
with $\Psi(\phi,z)=e^{i(m \phi + k_z z)}$.
For the $z$-components it is possible to choose the solutions
\begin{equation}
E_{z}^{(j)}=A^{(j)}f^{(j)}_{\mu}(q_{j} r)\Psi(\phi,z)\;;\;\;\;\;
H_z^{(j)}=B^{(j)} f^{(j)}_{\mu}(q_{j} r)\Psi(\phi,z)\;;\;\;(j=1,2)\;,
\label{fiedlz}
\end{equation}
where $q_{j}^{2}=\epsilon_i\omega^2/c^2-k_z^2$, $\epsilon_1 (\epsilon_2)$  is the inner  (outer) dielectric constant.
In Eq.~(\ref{fiedlz}) $f^{(1)}_{\mu}$ is the Bessel function $J_{\mu}(q_1 r)$ for $0\leq r\leq R$ and if $q_2^2>0$ ($q_2^2=-\kappa^2<0$),  $f^{(2)}_{m}$ is the Hankel function of the first kind, $\mathcal{H}_{\mu}^{(1)}(q_2 r)$ (MacDonald function $K_{\mu}(|q_2 |r)$)for $r\ge R$), both defined in Ref.~\onlinecite{Abramowitz}.

Using Eqs.~(\ref{amplitude}), the field transverse components are given by
\begin{equation}
\bigg[E_r^{(j)}(r),E_\phi^{(j)}(r)\bigg]=\frac{1}{q_i{^2}-k_z^2}\bigg[ - \frac{\omega \mu}{ c r}B^{(j)} f_{\mu}^{(j)} + ik_zq_jA^{(j)}f^{ '(j)}_{\mu}, - i \frac{\omega q_j}{c}B ^{(j)}f^{ '(j)}_{\mu}  - \frac{\mu k_z}{r}A^{(j)}f_{\mu}^{(j)} \bigg]\;;
\label{E}
\end{equation}
\begin{equation}
\bigg[H_r^{(j)}(r),H_\phi^{(j)}(r)\bigg]=\frac{1}{q_j^2-k_z^2}\bigg[ \frac{\epsilon_j\omega \mu}{c r}A ^{(j)}f_{\mu}^{(j)} + ik_zq_jB^{(j)} f^{ '(j)}_{\mu}, i \frac{\epsilon_j \omega q_j}{c}A^{(j)} f^{ '(j)}_{\mu} - \frac{\mu k_z}{r}B^{(j)} f_{\mu}^{(j)}\bigg]\;.
\label{H}
\end{equation}
It is possible now to determine the eigenmodes by imposing appropriate boundary conditions. The normal component of the electric induction vector, $\vec{\mathbb{D}}$, the electric field components  $[E_r,\; E_\phi]$ and the magnetic field vector,  $\vec{\mathbb{H}}$, should be continuous at the interface $r=R$, i.e. $\epsilon _1 E_r^{(1)}(R)=\epsilon _2 E_r^{(2)}(R)$, $E_{\phi}^{(1)} (R)= E_{\phi}^{(2)}(R)$, $E_z^{(1)} (R)= E_z^{(2)}(R)$ and $\vec{H}^{(1)}(R)=\vec{H}^{(2)}(R)$.
Using Eqs.~(\ref {E}) and~(\ref {H}) and assuming $q^2_j=-\kappa^2$, we obtain the dispersion relation
\begin{equation}
(q\kappa R)^2\big[\kappa K_{\mu}J'_{\mu}+q J_{\mu}K'_{\mu}\big]\big[\epsilon_1\kappa K_{\mu}J'_{\mu}+q\epsilon_2 J_{\mu}K'_{\mu}\big]=(\mu K_\mu J_\mu)^2\frac{\omega^2 k_z^2}{c^2}(\epsilon_1-\epsilon_2)^2\;.
\label{EcuSecular}
\end{equation}
For a given $\mu$ and $k_z$, Eq.~(\ref{EcuSecular}) provides a set of solutions, which we wish to compare to those of our "ideal cavity" model. An ideal cavity mode is one where the tangential components of the electric field and $H_r$ vanish at the interface. We can imagine that the walls of the ideal cavity are made of a perfect metal. Then $E_r$ is discontinuous at the interface and so are $H_z$ and $H_{\phi}$. Therefore, an ideal  cavity TM mode has the components that are non-zero only inside the cavity:
\begin{eqnarray}
\nonumber
E_{z_{1}}=C_{\mu}J_{\mu}(q r)\;,\\
E_{r_{1}}=  \frac{k_z}{q_1}C_{\mu} J_{\mu}'(q_1r)\;.
\label{34}
\end{eqnarray}
So, the boundary conditions reduce to $J_{\mu}(q_1 R)=0$. In Fig.~\ref{Fig:profiles} we compare the exact field profiles $E_z$ and $D_r$ with fields~(\ref{34}). As an example, it shows the $\nu =9$ mode, which corresponds to $\hbar \omega =$ 1.866 eV for ideal cavity of  $R=$  3$\mu m$ and demonstrates a reasonable qualitative agreement between the field distributions inside the cylinder. where the excitons are confined.

\begin{figure}[h]
\includegraphics[width=.5\linewidth]{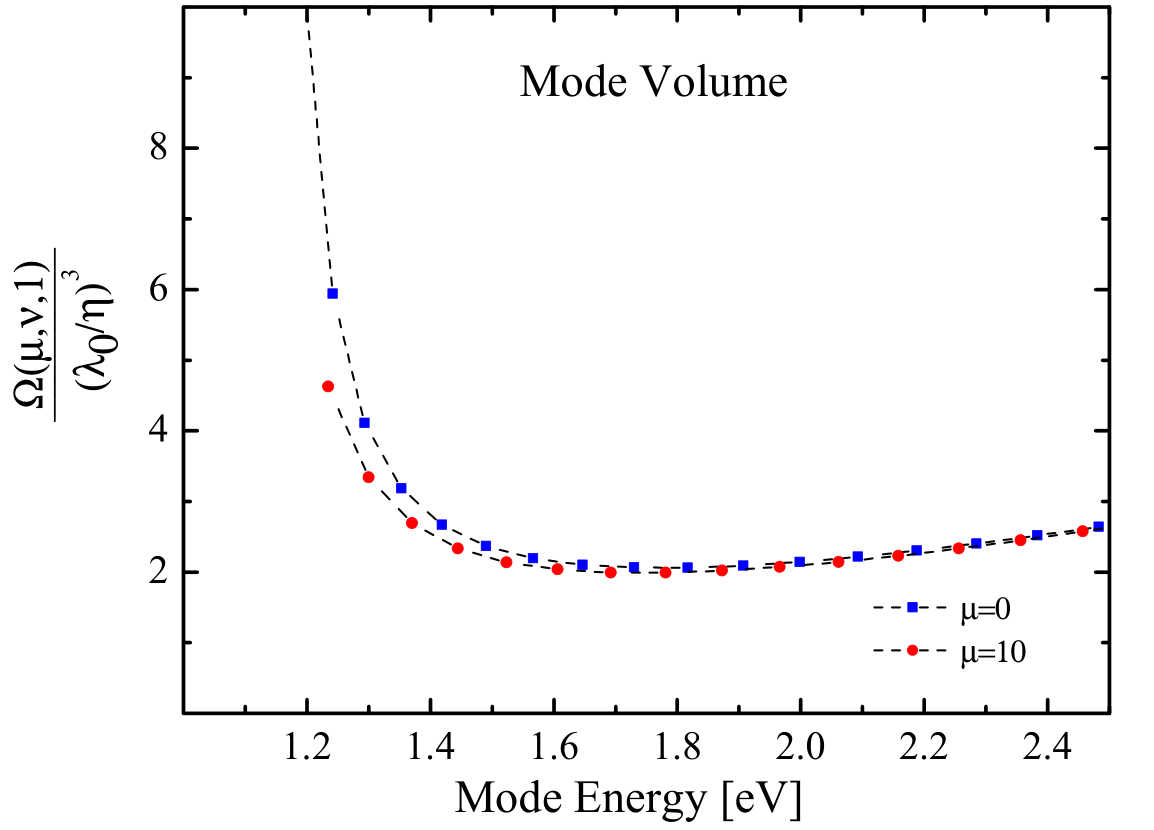}
\caption{mode volume {\it versus} mode energy calculated according to Eq. (\ref{modevolume2}) for ideal cavity with cylinder radius $R=3\;\mu$m and height $L=0.3\;\mu$m. $\lambda_0$ is the emission wavelength.}
\label{Fig:modevolume}
\end{figure}

We end this Appendix by presenting a plot of the mode volume dependence upon the mode energy, calculated according to Eq. (\ref {modevolume2}) for the ideal cavity (see Fig. \ref {Fig:modevolume}). 
%
%

\section{\label{appendix B} Exciton-photon matrix elements  $I_{\mu,\nu;m,n}$}

The integral $I_{\mu,\nu;m,n}$ in the matrix element (\ref {ME2}) is written as
 \begin{equation}
I_{\mu,\nu;m,n}=\frac 2 { \xi _{\mu,\nu}\vert J_{m}^\prime(\xi_{m,n})\vert}\int_0^1{xJ_{m}(\xi_{m,n}x)J_{m}(\xi_{\mu,\nu}x)dx}
\;.
\end{equation}
Using the Bessel functions' properties,~\cite{Abramowitz} the integral is equal to:
\begin{equation}
I_{\mu,\nu;m,n}=\frac{2}{\xi_{\mu,\nu}\vert J_m'(\xi_{m,n})\vert}\frac{\xi_{m,n}J_{m+1}(\xi_{m,n})J_m(\xi_{\mu,\nu})-\xi_{\mu,\nu}J_m(\xi_{m,n})J_{m+1)}(\xi_{\mu,\nu)}}{\xi_{m,n}^2-\xi_{\mu,\nu}^2},
\end{equation}
and, since the second term is zero, we obtain:
\begin{eqnarray}
\nonumber
I_{\mu,\nu;m,n}=\frac{2}{\xi_{\mu,\nu}\vert J_m'(\xi_{m,n})\vert}\frac{\xi_{m,n}J_{m+1}(\xi_{m,n})J_m(\xi_{\mu,\nu})}{\xi_{m,n}^2-\xi_{\mu,\nu}^2}\\
=-\frac{2}{\xi_{\mu,\nu}\vert J_m'(\xi_{m,n})\vert}\frac{\xi_{m,n}J'_m(\xi_{m,n})J_m(\xi_{\mu,\nu})}{\xi_{m,n}^2-\xi_{\mu,\nu}^2}.
\end{eqnarray}
Therefore, the integral squared is given by:
\begin{equation}
\vert I_{\mu,\nu;m,n} \vert^2=4\frac{\xi_{m,n}^2}{\xi_{\mu,\nu}^2}\frac{J^2_m(\xi_{\mu,\nu})}{(\xi_{m,n}^2-\xi_{\mu,\nu}^2)^2}\;.
\label {integrals}
\end{equation}
As an example, the squared values of the integrals  (\ref  {integrals}) for $\mu =0$ and $m=1$ are given in Table 1.
\begin{center}
%
%
\begin{table}[H]
\begin{center}
Table 1. Values of $\vert I_{0,\nu;1,n}\vert^2$.
\end{center}
\end{table}
\begin{tabular}{|p{0.5cm}|p{2cm}|p{2cm}|p{2cm}|p{2cm}|  }
\hline
$\nu$/n & 1 &2&3&4 \\
\hline
1 & $3.5 \times 10^{-2}$ & $4.8 \times 10^{-3}$ & $2.0\times 10^{-3}$ & $1.1\times 10^{-3}$ \\2 & $9.0 \times 10^{-4}$ & $2.1 \times 10^{-3}$ & $2.9\times 10^{-4}$ & $1.2\times 10^{-4}$ \\
3 & $1.6 \times 10^{-5}$ & $2.9 \times 10^{-4}$ & $5.0\times 10^{-4}$ & $6.6\times 10^{-5}$ \\
4 & $1.5 \times 10^{-6}$ & $9.5 \times 10^{-6}$ & $1.2\times 10^{-4}$ & $1.8\times 10^{-4}$\\
\hline
\end{tabular}
\end{center}


\bibliographystyle{apsrev}
\bibliography{Exciton-polaritons_in_cylindrical_cavity_v21}

\end{document}